\documentclass[10pt,journal,cspaper,compsoc]{IEEEtran}
\usepackage{paralist}

\usepackage{color}

\usepackage[ruled,linesnumbered]{algorithm2e}

\usepackage{adjustbox}
\usepackage{diagbox}

\usepackage{amsmath}
\usepackage{multirow}

\usepackage{subfigure}
\usepackage{graphicx}
\usepackage{epsfig}
\usepackage{epstopdf}

\usepackage{amssymb}
\usepackage{amsmath}

\usepackage{bm}
\usepackage{booktabs}
\usepackage{verbatim}
\usepackage{array}
\usepackage{hyperref} 

\usepackage{balance}

\DeclareMathOperator*{\argmax}{argmax}

\begin{document}
\title{Learning from Atypical Behavior: Temporary Interest Aware Recommendation Based on Reinforcement Learning}

\author{Ziwen~Du,~
        Ning~Yang,~\IEEEmembership{Member,~IEEE,}
        Zhonghua Yu,~
        Philip S.~Yu,~\IEEEmembership{Fellow,~IEEE}
        % <-this % stops a space
\IEEEcompsocitemizethanks{
\IEEEcompsocthanksitem Ziwen Du is with the School
of Computer Science, Sichuan University, China. \protect 
 E-mail: ddzzww123456@gmail.com
 
\IEEEcompsocthanksitem Ning Yang is the corresponding author and with the School
of Computer Science, Sichuan University, China. \protect 
 E-mail: yangning@scu.edu.cn
 
 \IEEEcompsocthanksitem Zhonghua Yu is with the School
of Computer Science, Sichuan University, China. \protect 
 E-mail: yuzhonghua@scu.edu.cn

\IEEEcompsocthanksitem Philip S. Yu is with the Department of Computer Science, University of Illinois at Chicago, USA. \protect 
E-mail: psyu@uic.edu
}% <-this % stops a space
\thanks{}}

%\markboth{Journal of \LaTeX\ Class Files,~Vol.~6, No.~1, January~2007}%
%{Shell \MakeLowercase{\textit{et al.}}: Bare Demo of IEEEtran.cls for Computer Society Journals}

\IEEEcompsoctitleabstractindextext{%
\begin{abstract}

Traditional robust recommendation methods view atypical user-item interactions as noise and aim to reduce their impact with some kind of noise filtering technique, which often suffers from two challenges. First, in real world, atypical interactions may signal users' temporary interest different from their general preference. Therefore, simply filtering out the atypical interactions as noise may be inappropriate and degrade the personalization of recommendations. Second, it is hard to acquire the temporary interest since there are no explicit supervision signals to indicate whether an interaction is atypical or not. To address this challenges, we propose a novel model called Temporary Interest Aware Recommendation (TIARec), which can distinguish atypical interactions from normal ones without supervision and capture the temporary interest as well as the general preference of users. Particularly, we propose a reinforcement learning framework containing a recommender agent and an auxiliary classifier agent, which are jointly trained with the objective of maximizing the cumulative return of the recommendations made by the recommender agent. During the joint training process, the classifier agent can judge whether the interaction with an item recommended by the recommender agent is atypical, and the knowledge about learning temporary interest from atypical interactions can be transferred to the recommender agent, which makes the recommender agent able to alone make recommendations that balance the general preference and temporary interest of users. At last, the experiments conducted on real world datasets verify the effectiveness of TIARec.

\end{abstract}

\begin{keywords}
Robust Recommendation, Temporary Interest, Deep Reinforcement Learning
\end{keywords}}

\maketitle

\IEEEdisplaynotcompsoctitleabstractindextext

\IEEEpeerreviewmaketitle

\section{Introduction}
In recent years, recommender systems have received much attention due to their ability to improve user experience in online applications like Amazon, Netflix, TikTok, etc. By nature, a recommender system makes recommendations based on the preference learned from users' historical interactions with items. However, in reality there always exist atypical behaviors that seldom appear before and obviously deviate from the general preference revealed by users' usual behaviors. For example, a user who is interested in football may occasionally watch some basketball-related videos during the finals of NBA, and when the NBA season is over, her attention to football will come back. The traditional works often consider the atypical interactions as noise that will impede the learning of the true preference of users and consequently weaken the robustness of recommender systems \cite{Deldjoo2020}.

Recently, researchers have proposed a few models for robust recommendation, whose main idea is to reduce the impact of atypical interactions with some kind of noise filtering technique \cite{Vincent2008Extracting,Yao2016Collaborative,Tang2018Adversarial,Yuan2019Adversarial,Zhang2020}. Simply filtering out atypical interactions, however, may be inappropriate for recommender systems in real world. On the one hand, due to the dynamic nature of user preference, even so called atypical interactions do not have to mean nothing but can also reveal users' temporary interest. Consider the above NBA example again. Although the behaviors of watching basketball-related videos seldom happen before, they can still be a signal of the user's temporary interest in basketball due to NBA. Therefore, simply removing atypical interactions as noise may degrade the learning of user preference, especially the learning of user temporary interest. On the other hand, it is hard for the existing recommendation models to acquire the temporary interest directly from users' historical interactions, since there are no explicit supervision signals to indicate atypical interactions \cite{zhang2019hierarchical}. 

\begin{figure}[t]
\centering
\subfigure[]{
\begin{minipage}[t]{0.20\textwidth}
\centering
\includegraphics[scale=0.40]{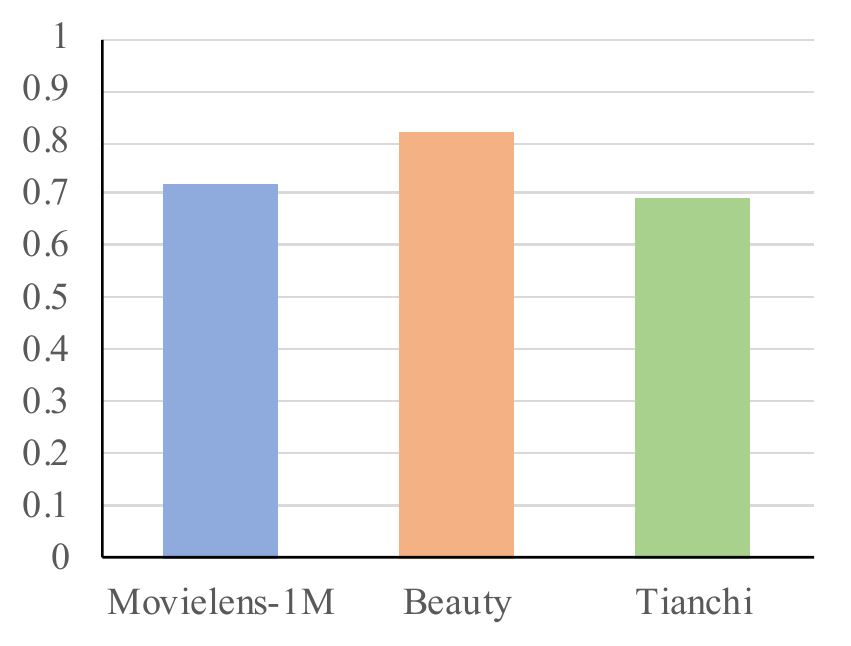}
\label{fig:statistics}
\end{minipage}%
}
\subfigure[]{
\begin{minipage}[t]{0.25\textwidth}
\centering
\includegraphics[scale=0.39]{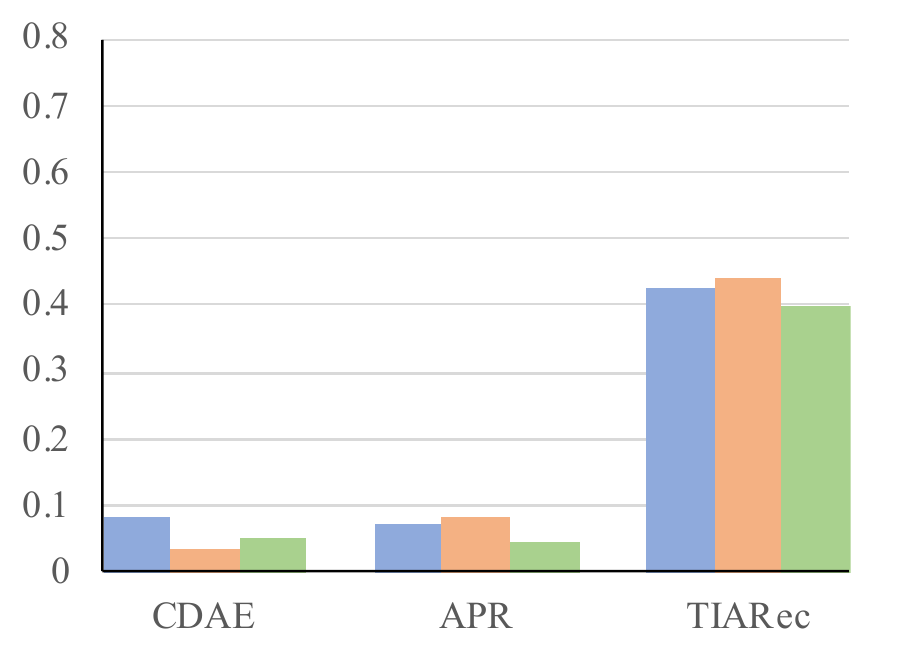}
\label{fig:recall}
\end{minipage}
}
\caption{(a) Statistics of the users with rare interactions in real datasets Movielens-1M, Beauty, and Tianchi. (b) The recalls of the users with atypical interactions provided by the two state-of-the-art robust recommendation models CDAE, APR, and the proposed model TIARec.}
\label{fig:data}
\end{figure}

To further illustrate our motivation, we conduct an exploration of three real datasets Movielens-1M, Beauty, and Tianchi, where the interactions are labeled in terms of the item category. Figure \ref{fig:statistics} shows the proportions of the users who have at least one rare interaction label with frequency less than 5\% of the average, while Figure \ref{fig:recall} shows the ratios of the users whose at least one rare interaction is predicted correctly by the robust recommendation models CDAE \cite{CDAE}  and APR \cite{APR} in addition to the proposed model TIARec on each dataset. We can see that although there are more than 70\% users with atypical behaviors in each dataset, the temporary preferences are discovered for no more than 10\% of them by the existing two robust recommendation models CDAE and APR. Therefore, we need a new method that is able to distinguish whether an interaction is atypical or not and capture the temporary interest from atypical interactions. 

To address the above challenges, motivated by the success of the reinforcement learning (RL) based recommendation models \cite{DEERs,DQN,zou2019reinforcement,shang2019environment,feng2018learning,ie2019slateq,he2020learning}, we propose a novel RL-based model, called Temporary Interest Aware Recommendation (\textbf{TIARec}), which is able to distinguish atypical interactions from normal ones without annotations and capture the temporary interest as well as the general preference. TIARec contains a recommender agent and an auxiliary classifier agent. The recommender agent recommends an item to a user as its action at each time step, while the classifier agent can help to identify whether the interaction with the recommended item is atypical in a trial-error fashion. We also introduce a critic network and employ an Actor-Critic algorithm \cite{Lowe2017} to jointly train the recommender agent and the classifier agent with an objective to maximize the expected cumulative return of the recommendations made by the recommender agent. During the joint training, the knowledge about how to capture the temporary interest can be transferred from the classifier agent to the recommender agent. Once the training is finished, the recommender agent can alone make recommendations that balance the general preference and temporary interest of users.

Our contributions can be summarized as follows:
\begin{itemize}
\item We propose a novel RL-based model for robust recommendation, called Temporary Interest Aware Recommendation (\textbf{TIARec}). To our best knowledge, TIARec is the first reinforcement learning based solution to capture temporary interests for robust recommendation.

\item  We propose a deep reinforcement learning network including a recommender agent and an auxiliary classifier agent. During their joint training, the auxiliary classifier agent teaches the recommender agent the knowledge about learning the temporary interest from atypical interactions.

\item Extensive experiments conducted on real world datasets verify the superiority of TIARec over the state-of-the-art methods on robust recommendation and RL-based recommendation. 
\end{itemize}

The rest of this paper is organized as follows. In Section 2, we introduce the preliminaries and formally define the target problem. In Section 3, we first give an overview of TIARec and then describe its details. We empirically evaluate the performance of TIARec over real-world datasets in Section 4. At last, we briefly review the related works in Section 5 and conclude in Section 6.

\section{Preliminaries}
In this section, we first give some basic definitions, then formulate the recommendations as a Markov Decision Process (MDP), and formally define the target problem of this paper. 

\subsection{Basic Definitions and Vectorization}
Let $\mathcal{U}$ be the user set and $\mathcal{V}$ be the item set. Let $\mathcal{O} = \{ o_1$, $\cdots$, $o_{|\mathcal{O}|} \}$ be an interaction sequence, where $o_i \in \mathcal{V}$, $1 \le i \le |\mathcal{O}|$, is the item of the $i$th interaction. Let $\mathcal{O}_t$, $\mathcal{M}_t$, and $\mathcal{N}_t$ be the sets of all interactions, normal interactions, and atypical interactions of some user that happen before time $t$, respectively. For the same user, $\mathcal{M}_t \cap \mathcal{N}_t = \Phi$ and $\mathcal{O}_t = \mathcal{M}_t \cup \mathcal{N}_t$. 
For vectorizing, we pretrain a $d$-dimensional vector for each item using Skip-gram algorithm, where an item $o_i$ is treated as a word and a historical interaction sequence $\mathcal{O}_t$ as a sentence. For simplicity, in the rest of this paper we use the pretrained embeddings to represent the corresponding items, which are denoted by bold lowercase, for example $\boldsymbol{o}_i$, and reuse $\mathcal{V}$ to denote the whole set of the item embeddings. 

\subsection{MDP Formulation}
Now we formulate the recommendations as an MDP, which is formally defined as follows:
\begin{itemize}

\item \textbf{State space} $\mathcal{S}$ In recommender systems, the environment of the agents are defined as the users. Therefore the environment state is defined by a user's interaction history. Particularly, a state $s$ is defined as any possible interaction set $\mathcal{O}$, i.e., the state space $\mathcal{S} = \{ s = \mathcal{O} \}$.  

\item \textbf{Action space} $\mathcal{A}$ An action $\boldsymbol{a}$ of a recommender system is defined as an embedding of a recommended item, and therefore the action space  $\mathcal{A}= \mathcal{V}$. In this paper we assume only one item will be recommended at each time step.

\item \textbf{Transitions} $\mathcal{P}$ The transition $p(s' | s, \boldsymbol{a})$ is the conditional probability that the state of the environment (user) changes from $s \in \mathcal{S}$ to $s' \in \mathcal{S}$ after an action $\boldsymbol{a} \in \mathcal{A} $ is issued. In recommender systems, $p(s' | s, \boldsymbol{a})$ is 1 if the user interacts with the recommended item, otherwise 0.

\item \textbf{Reward} $\mathcal{R}$ For a recommender, the immediate reward $r(s,\boldsymbol{a})$ after issuing action $\boldsymbol{a}$ at state $s$ will be defined in terms of users' feedbacks. 

\item \textbf{Discount rate} $\gamma$ $\gamma \in (0,1] $ is the discount factor for measuring long-term rewards.

\end{itemize}

\subsection{Problem Formulation}
Let $\pi : \mathcal{S} \times \mathcal{V} \mapsto [0,1]$ be a recommender policy that outputs the conditional probability $p(\boldsymbol{a}|s)$ of recommending item $\boldsymbol{a} \in \mathcal{V}$ at state $s \in \mathcal{S}$. Let $\mathcal{V}^u = \{ \boldsymbol{v}^u_1, \cdots, \boldsymbol{v}^u_T\}$ be a sequence of the collected historical interactions of some user $u$ that really happen before time step $T$, where $T$ is the number of time steps considered and $\boldsymbol{v}^u_i \in \mathcal{V}$ ($1 \le i \le T$) is the item embedding of the $i$th interaction. Given the dataset $\{ \mathcal{V}^u \}$ over all users, we want to learn an optimal policy $\pi^*$ for the recommender agent that maximizes the expected cumulative reward of the recommender, i.e., 
\begin{equation}
\pi^* = \argmax_{\pi}\mathbb{E} \Big[ \sum_{t = 0}^{T} \gamma^{t} r(s_t, \boldsymbol{a}_t) \Big],
\label{eq:target}
\end{equation}
where $s_t = {\mathcal{O}_t} \in \mathcal{S}$ and $\boldsymbol{a}_t \in \mathcal{V}$ are the user state and the recommended item at time step $t$, respectively.

\section{The Proposed Model}
In this section, we first give an overview of the architecture of TIARec, and then describe its components and learning in detail.

\begin{figure}[!t]
  \centering
  \includegraphics[width=0.35\textwidth]{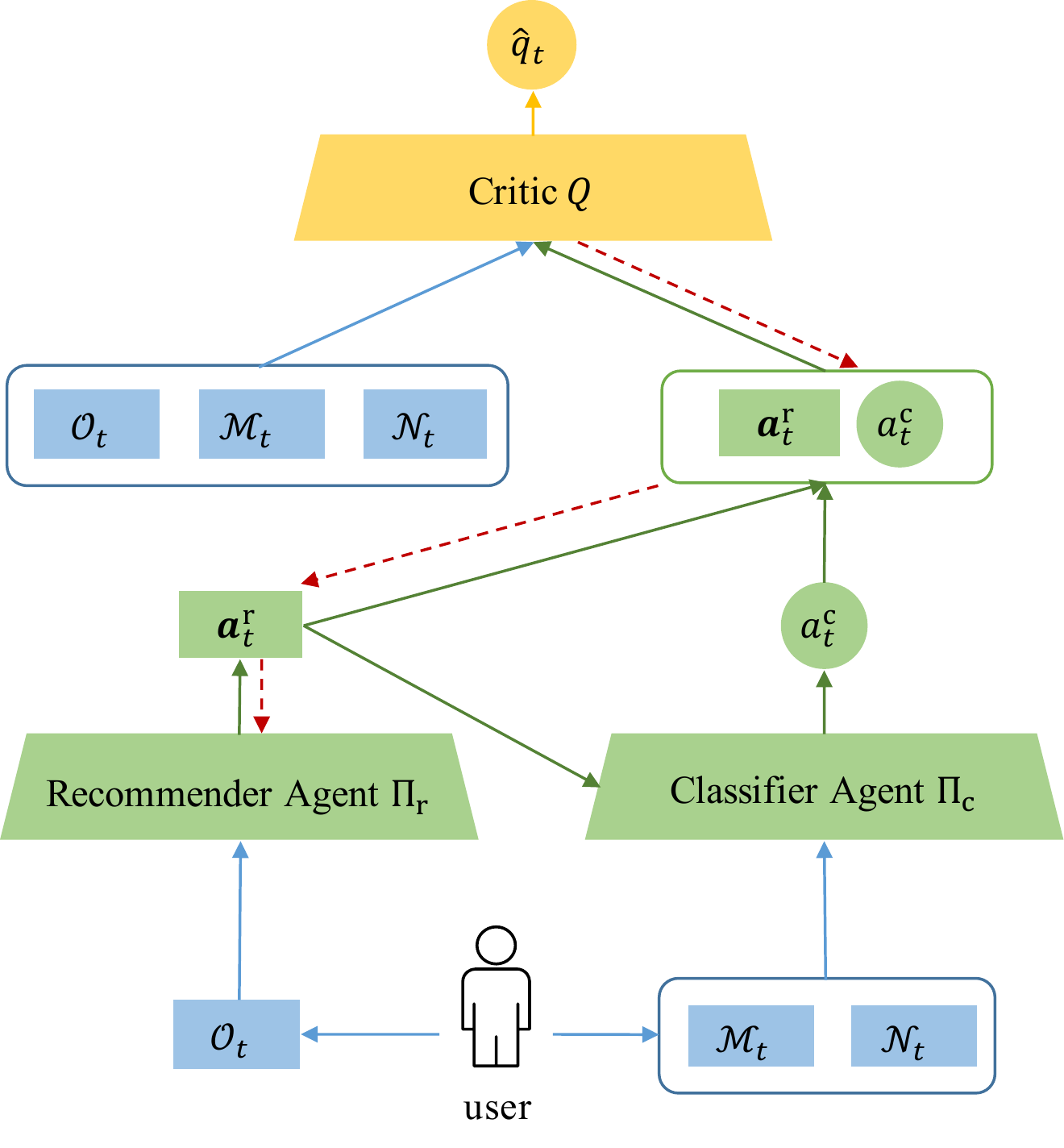}\\
  \caption{The architecture of TIARec}
  \label{fig:overmodel}
\end{figure}

\subsection{Architecture of TIARec}

As shown in Figure \ref{fig:overmodel}, TIARec consists of three parts, a recommender agent $\Pi_{\text{r}}$, an auxiliary classifier agent $\Pi_{\text{c}}$, and a critic $Q$. The goal of TIARec is to learn an optimal recommender agent $\Pi_{\text{r}}$ that realizes the objective defined by Equation (\ref{eq:target}), with the help of the classifier agent and the critic. 

The agents $\Pi_{\text{r}}$ and $\Pi_{\text{c}}$ are two policy networks. At each time step $t$, based on the current state $s^{\text{r}}_t = \mathcal{O}_t$, the recommender agent recommends an item 
\begin{equation}
\boldsymbol{a}^{\text{r}}_t  = \Pi_{\text{r}}(s^{\text{r}}_t; \Theta_{\text{r}}) \in \mathcal{V}
\end{equation}
as the action, where $\boldsymbol{a}^{\text{r}}_t \in \mathbb{R}^{d \times 1}$ is an item embedding and $ \Theta_{\text{r}}$ represents the trainable parameters of $\Pi_{\text{r}}$. Then the classifier agent judges how likely the interaction with $\boldsymbol{a}^{\text{r}}_t$ is atypical. The state $s^{\text{c}}_t$ observed by the classifier agent includes three parts, the action $\boldsymbol{a}^{\text{r}}_t$ of the recommender agent, the current normal interaction set $\mathcal{M}_t$ and atypical interaction set $\mathcal{N}_t$. Based on its current state $s^{\text{c}}_t = \{ \boldsymbol{a}^{\text{r}}_t,\mathcal{M}_t, \mathcal{N}_t \}$, the classifier agent outputs a scalar 
\begin{equation}
a^{\text{c}}_t= \Pi_{\text{c}} \big(s^{\text{c}}_t ; \Theta_{\text{c}} \big) \in [0,1],
\end{equation}
where $\Theta_{\text{c}}$ is the trainable parameters of $\Pi_{\text{c}}$. The real number $a^{\text{c}}_t$ is a probability defining the action of the classifier agent, which is to recognize the interaction with $\boldsymbol{a}^{\text{r}}_t$ as atypical and append it to the atypical interaction set $\mathcal{N}_t$ with probability $a^{\text{c}}_t$, or as a normal interaction and append it to the normal interaction set $\mathcal{M}_t$ with probability $1-a^{\text{c}}_t$. Essentially, $\mathcal{N}_t$ encodes a user's temporary interest while $\mathcal{M}_t$ encodes the user's general preference.

The critic network $Q$ plays the role of the state-action value function in Actor-Critic algorithm, which helps the recommender agent and the classifier agent understand how good the states and the actions are in terms of the expected cumulative return. With the overall state $s^{\text{q}}_t = \{\mathcal{O}_t, \mathcal{M}_t, \mathcal{N}_t \}$ and the overall action $\boldsymbol{a}^{\text{q}}_t = \boldsymbol{a}^{\text{r}}_t \oplus a^{\text{c}}_t$ as input, where $\oplus$ represents concat operation and $\boldsymbol{a}^{\text{q}}_t \in \mathbb{R}^{(d+1) \times 1}$, the critic evaluates the expected cumulative return $\widehat{q}_t$ for both the recommender agent and the classifier agent,
\begin{equation}
\widehat{q}_t = Q \big(s^{\text{q}}_t, \boldsymbol{a}^{\text{q}}_t; \Theta_{\text{q}} \big),
\end{equation}
where $\Theta_{\text{q}}$ represents the trainable parameters of the critic. The overall objective of the training is to learn the optimal $\Theta_{\text{r}}$, $\Theta_{\text{c}}$, and $\Theta_{\text{q}}$ that maximize the expected cumulative return $\widehat{q}_t$.

The insight here is that the classifier agent $\Pi_\text{c}$ can help capture the temporary interest carried by the atypical interactions it recognizes, due to the trial-error nature of reinforcement learning. As the two agents are jointly trained, the knowledge about learning temporary interest from atypical interactions can be transferred from the classifier agent to the recommender agent along the back propagation path marked by the red dashed lines in Figure \ref{fig:overmodel}. For example, if the classifier agent correctly recognizes an interaction as atypical (or mistakes it as a normal one), then the cumulative return will consequently increases (or reduces), and during the joint training, the supervision signal offered by the critic network will not only guide the update of the classifier agent, but also enforce the parameters of the recommender agent to be adjusted such that the recommender agent can remember that in such case, it should recommend an appropriate item satisfying the user's temporary interest rather than her general preference, so as to maximize the cumulative return.

\subsection{Recommender Agent $\Pi_{\text{r}}$}
At each time step $t$, the task of the recommender agent $\Pi_{\text{r}}(s^{\text{r}}_t; \Theta_{\text{r}})$ is to predict an item $\boldsymbol{a}^{\text{r}}_t$ that a given user will interact with next time according to the current state $s_t = \mathcal{O}_t$ of that user, which is implemented as an MLP combining an attention network. Particularly, we first generate the current preference embedding $\boldsymbol{z}^{\text{o}}_t \in \mathbb{R}^{d\times1}$ based on the user's current state with following attention network:
\begin{equation}
\begin{split}
\boldsymbol{z}^{\text{o}}_t &= \sum_{i=1}^{|\mathcal{O}_t|}\alpha^{\text{o}}_i \boldsymbol{o}_i, \\
\alpha^{\text{o}}_i &= \frac{\exp \big(\boldsymbol{q}^\text{T} \sigma(\boldsymbol{W}^{\text{o}}\boldsymbol{o}_i + \boldsymbol{b}^{\text{o}})  \big)}{ \sum_{j=1}^{|\mathcal{O}_t|} \exp \big( \boldsymbol{q}^\text{T} \sigma(\boldsymbol{W}^{\text{o}}\boldsymbol{o}_j + \boldsymbol{b}^{\text{o}}) \big)}
\end{split}
\label{eq:zo}
\end{equation}
where $\boldsymbol{o}_i \in \mathbb{R}^{d\times1}$ represents the pretrained embedding of the $i$th item $o_i \in \mathcal{O}_t$, $\alpha^{\text{o}}_i \in \mathbb{R}$ is the attention coefficient, and $\boldsymbol{q} \in \mathbb{R}^{d\times1}$, $\boldsymbol{W}^{\text{o}} \in \mathbb{R}^{d \times d}$, and $\boldsymbol{b}^{\text{o}} \in \mathbb{R}^{d \times 1}$ are the trainable query vector, transformation matrix, and bias terms, respectively, and $\sigma(\cdot)$ is a non-linear activation function.

Then we feed $\boldsymbol{z}^{\text{o}}_t$ into an MLP network to generate a prototype item embedding $\boldsymbol{a}^{\text{p}}_t \in \mathbb{R}^{d\times1}$, 
\begin{equation}
\boldsymbol{a}^{\text{p}}_t = \text{MLP}_\text{r} (\boldsymbol{z}^{\text{o}}_t; \theta_{\text{r}}),
\label{eq:prototype}
\end{equation}
where $\theta_{\text{r}}$ represents the trainable parameters of $\text{MLP}_\text{r}$. Now the set of the trainable parameters of the recommender agent is $ \Theta_{\text{r}} = \{ \theta_{\text{r}},\boldsymbol{W}^{\text{o}},  \boldsymbol{b}^{\text{o}}, \boldsymbol{q} \}$.

Note that $\boldsymbol{a}^{\text{p}}_t$ is an embedding of the \textbf{\textit{prototype item}} that may not exist in $\mathcal{V}$ but defines how the embedding of the best recommended item should look. To generate a real item as the action and set its immediate reward, we generate a top-$k$ list $\mathcal{V}_k$ including the first $k$ closest item embeddings of $\boldsymbol{a}^{\text{p}}_t$ in terms of cosine similarity. The item whose embedding is closest to $\boldsymbol{a}^{\text{p}}_t$ is chosen as the action of the recommender agent during the training,
\begin{equation}
\boldsymbol{a}^{\text{r}}_t = \argmax_{\boldsymbol{v} \in \mathcal{V}} \cos(\boldsymbol{a}^{\text{p}}_t, \boldsymbol{v}).
\label{eq:r_action}
\end{equation}
where $\cos(\cdot , \cdot)$ is the cosine similarity and $\boldsymbol{v} \in \mathbb{R}^{d\times1}$ is the pretrained embedding of item $v$. 

For offline training, we need to obtain the immediate reward $R^{\text{r}}_t$ of the action $\boldsymbol{a}^{\text{r}}_t$ at time step $t$. The challenge here is that the logged data $\{ \mathcal{V}^{u} \}$ were generated by an uncontrollable recommender system whose recommending policy is unknown, or in other words, the historical data are user feedbacks to another unknown agent different from ours. Therefore, during the offline training, we cannot directly set $R^{\text{r}}_t$ with respect to whether or not the recommended list $\mathcal{V}_k$ covers the logged item that the user actually interacts with at time step $t$. The existing works often apply off-policy strategy to fulfill the offline training, by which the target recommending policy can be updated according to the data generated by a different but known policy \cite{chen2019top-k,Zhao2018}. However, the traditional offline training strategy is inapplicable to our case, since in our case, the policy generating the logged data is unknown. Inspired by the techniques used in \cite{chen2018large-scale,DDPG,zou2020pseudo,Liu2020}, our idea to overcome this dilemma is to mimic the online interactions of the recommender agent with the user by inferring how likely the user will interact with an item in the recommended list based on the logged offline data. In particular, before the training, we first learn a user latent factor $\boldsymbol{u} \in \mathbb{R}^{d \times 1}$ for each user $u \in \mathcal{U}$ and an item latent factor $\boldsymbol{v} \in \mathbb{R}^{d \times 1}$ for each item $v \in \mathcal{V}$, by performing probabilistic matrix factorization (PMF) over the whole dataset in advance. Then the probability of user $u$ will interact with item $v$ can be calculated as 
\begin{equation}
p(\boldsymbol{u}, \boldsymbol{v}) = \sigma(\boldsymbol{u}^{\text{T}} \boldsymbol{v}),
\end{equation}
where $ \sigma(\cdot)$ is sigmoid function. Next, during the training, we set the immediate reward $R^{\text{r}}_t$ to the inferred probability that user $u$ will interact with at least one item in the recommended list $\mathcal{V}_k$, i.e., 
\begin{equation}
R^{\text{r}}_t = \frac{1}{k}\sum_{\boldsymbol{v} \in \mathcal{V}_k} p(\boldsymbol{u}, \boldsymbol{v}).
\label{eq:r_reward}
\end{equation}
The insight here is that the larger the probability that user will interact with the recommended list, the higher the quality of it, and the more reward the agent deserves.

\subsection{Classifier Agent $\Pi_{\text{c}}$}

The task of the classifier agent $\Pi_{\text{c}} \big(s^{\text{c}}_t = \{ \boldsymbol{a}^{\text{r}}_t,\mathcal{M}_t, \mathcal{N}_t \}; \Theta_{\text{c}} \big)$ is to determine how likely an interaction with the item $\boldsymbol{a}^{\text{r}}_t$ recommended by the recommender agent is atypical, based on current $\{ \mathcal{M}_t, \mathcal{N}_t \}$. Similar to the structure of the recommender agent, the classifier agent is also implemented as an MLP but combining two attention networks. 

At each time step $t$, to capture the temporary interest from the atypical interactions, the classifier agent first generates a temporary interest embedding $\boldsymbol{z}^{\text{n}}_t \in \mathbb{R}^{d \times 1}$ based on the current atypical interaction set $\mathcal{N}_t$, with the following attention network: %similar to Equation (\ref{eq:zo}):
\begin{equation}
\begin{split}
\boldsymbol{z}^{\text{n}}_t &= \sum_{i=1}^{|\mathcal{N}_t|}\alpha^{\text{n}}_i \boldsymbol{n}_i, \\
\alpha^{\text{n}}_i &= \frac{\exp \big[ (\boldsymbol{a}^{\text{r}}_t)^\text{T} \sigma(\boldsymbol{W}^{\text{n}}\boldsymbol{n}_i + \boldsymbol{b}^{\text{n}}) \big]}{ \sum_{j=1}^{|\mathcal{N}_t|} \exp \big[ (\boldsymbol{a}^{\text{r}}_t)^\text{T} \sigma(\boldsymbol{W}^{\text{n}}\boldsymbol{n}_j + \boldsymbol{b}^{\text{n}}) \big]},
\end{split}
\label{eq:zn}
\end{equation}
where $\boldsymbol{n}_i \in \mathbb{R}^{d\times1}$ represents the pretrained embedding of the $i$th item in $\mathcal{N}_t$, $\alpha^{\text{n}}_i \in \mathbb{R}$ is the attention coefficient, $\boldsymbol{W}^{\text{n}} \in \mathbb{R}^{d \times d}$ and $\boldsymbol{b}^{\text{n}} \in \mathbb{R}^{d \times 1}$ are the learnable transformation matrix and bias terms, respectively, and $\boldsymbol{a}^{\text{r}}_t$ serves as the query vector. Similarly, the classifier agent also generate a general preference embedding $\boldsymbol{z}^{\text{m}}_t \in \mathbb{R}^{d \times 1}$ based on the current normal interaction set $\mathcal{M}_t$,
\begin{equation}
\begin{split}
\boldsymbol{z}^{\text{m}}_t &= \sum_{i=1}^{|\mathcal{M}_t|}\alpha^{\text{m}}_i \boldsymbol{m}_i, \\
\alpha^{\text{m}}_i &= \frac{\exp \big[ (\boldsymbol{a}^{\text{r}}_t)^\text{T} \sigma(\boldsymbol{W}^{\text{m}}\boldsymbol{m}_i + \boldsymbol{b}^{\text{m}}) \big]}{ \sum_{j=1}^{|\mathcal{M}_t|} \exp \big[ (\boldsymbol{a}^{\text{r}}_t)^\text{T} \sigma(\boldsymbol{W}^{\text{m}}\boldsymbol{m}_j + \boldsymbol{b}^{\text{m}}) \big]}.
\end{split}
\label{eq:zm}
\end{equation}
Finally, the classifier agent feeds the concatenation of the embeddings $\boldsymbol{z}^{\text{n}}_t $ and $\boldsymbol{z}^{\text{m}}_t $ into an MLP to generate the probability $a^{\text{c}}_t$ that $\boldsymbol{a}^{\text{r}}_t$ is an atypical interaction,
\begin{equation}
\begin{split}
a^{\text{c}}_t 
= \text{MLP}_{\text{c}} (\boldsymbol{z}^{\text{n}}_t \oplus \boldsymbol{z}^{\text{m}}_t; \theta_{\text{c}}) \in [0,1],
\end{split}
\label{eq:c_action}
\end{equation}
where $\oplus$ is the concatenation operation and $\theta_{\text{c}}$ represents the trainable parameters of $\text{MLP}_{\text{c}}$. Now we can see that the trainable parameter set of the classifier agent is $\Theta_{\text{c}} = \{ \theta_{\text{c}}, \boldsymbol{W}^{\text{n}}, \boldsymbol{W}^{\text{m}}, \boldsymbol{b}^{\text{n}}, \boldsymbol{b}^{\text{m}} \}$.

As we mentioned before, the probability $a^{\text{c}}_t$ actually defines two alternative actions for the classifier agent, where one is to recognize $\boldsymbol{a}^{\text{r}}_t$ as atypical with probability $a^{\text{c}}_t$ and append it to $\mathcal{N}_t$, i.e.,
\begin{equation}
\mathcal{N}_{t+1} = \mathcal{N}_t \cup \{ \boldsymbol{a}^{\text{r}}_t \},
\end{equation} 
and the other one is to recognize $\boldsymbol{a}^{\text{r}}_t$ as a normal interaction with probability $1- a^{\text{c}}_t$ and append it to $\mathcal{M}_t$, i.e.,
\begin{equation}
\mathcal{M}_{t+1} = \mathcal{M}_t \cup \{ \boldsymbol{a}^{\text{r}}_t \}.
\end{equation} 
As we will see later, at the beginning of a training cycle, $\mathcal{N}_0$ and $\mathcal{M}_0$ are initialized with empty set, and they will gradually grow with more actions issued by the recommender agent.

The classifier agent also needs to obtain an immediate reward for its action. Intuitively, atypical behaviors should be dissimilar to each other and lead to diversity, otherwise many similar atypical behaviors will not be atypical any more. Therefore, if $\boldsymbol{a}^{\text{r}}_t$ is recognized as atypical appropriately, the overall similarity between the items in $\mathcal{N}_t$ should reduce after $\boldsymbol{a}^{\text{r}}_t$ is merged into $\mathcal{N}_t$, which implies that smaller similarity between $\boldsymbol{a}^{\text{r}}_t$ and the items in $\mathcal{N}_t$ should receive greater reward. On the contrary, if $\boldsymbol{a}^{\text{r}}_t$ is recognized as normal appropriately, the overall similarity between the items in $\mathcal{M}_t$ should rise after $\boldsymbol{a}^{\text{r}}_t$ is merged into $\mathcal{M}_t$, which implies that larger similarity between $\boldsymbol{a}^{\text{r}}_t$ and the items in $\mathcal{M}_t$ is better. Based on this idea, we set the immediate reward $R^{\text{c}}_t$ for the classifier agent as
\begin{equation}
R^{\text{c}}_t = 
\begin{cases}
1- \frac{1}{|\mathcal{N}_t|} \sum_{i}^{|\mathcal{N}_t|} \cos(\boldsymbol{n}_i, \boldsymbol{a}^{\text{r}}_t), \text{  if } a^{\text{r}}_t \in \mathcal{N}_{t+1}, \\ 
\\

\frac{1}{|\mathcal{M}_t|} \sum_{i}^{|\mathcal{M}_t|} \cos(\boldsymbol{m}_i, \boldsymbol{a}^{\text{r}}_t), \text{  if } a^{\text{r}}_t \in \mathcal{M}_{t+1}.
\end{cases}
\label{eq:c_reward}
\end{equation}

There are two points here we want to further clarify. First, as there are no supervision signals, how can the classifier agent learn to recognize atypical behavior? Our idea here is that atypical behaviors cause dissimilarity, and the immediate reward given by Equation (\ref{eq:c_reward}) and the expected cumulative return estimated by the critic (see next subsection) will judge how well the classifier agent classifies a recommended item as atypical or normal, and accordingly adjust its parameters through the policy gradient descent during the training process. Second, where is the temporary interest? As users' temporary interest hides in their atypical interactions, we argue that the atypical interaction set $\mathcal{N}_t$ reveals what the classifier agent considers the temporary interest, and during the joint training, such knowledge will be transferred to the recommender agent. 

\subsection{Critic $Q$}
The critic network $Q (s^{\text{q}}_t, \boldsymbol{a}^{\text{q}}_t; \Theta_{\text{q}} )$ is responsible for evaluating the expected cumulative return $\widehat{q}_t$ based on the state $s^{\text{q}}_t = \{\mathcal{O}_t, \mathcal{M}_t, \mathcal{N}_t \}$ and action $\boldsymbol{a}^{\text{q}}_t = \boldsymbol{a}^{\text{r}}_t \oplus a^{\text{c}}_t$. The critic generates the estimation of the expected cumulative return through an MLP,
\begin{equation}
\widehat{q}_t = \text{MLP}_{\text{q}}(\boldsymbol{a}^{\text{q}}_t \oplus \boldsymbol{z}^{\text{o}}_t \oplus \boldsymbol{z}^{\text{n}}_t \oplus \boldsymbol{z}^{\text{m}}_t ; \Theta_{\text{q}}),
\label{eq:critic}
\end{equation}
where $\boldsymbol{z}^{\text{o}}_t$, $\boldsymbol{z}^{\text{n}}_t$, and $\boldsymbol{z}^{\text{m}}_t$ are the preference embeddings of $\mathcal{O}_t$, $\mathcal{M}_t$,  and $\mathcal{N}_t$, respectively, which are produced by Equations (\ref{eq:zo}), (\ref{eq:zn}), and  (\ref{eq:zm}), respectively.

\subsection{Learning of TIARec}
As we have mentioned, the recommender agent is trained jointly with the classifier agent and the critic network using an actor-critic algorithm. In particular, we employ MADDPG \cite{Lowe2017} algorithm to learn the parameters $\Theta_{\text{r}}$, $\Theta_{\text{c}}$, and $\Theta_{\text{q}}$ of TIARec.

\subsubsection{Target Networks}
The target networks are used to provide the supervision signal for the training. The target networks $\Pi'_{\text{r}}(s^{\text{r}}_{t+1}; \Theta'_{\text{r}})$ and $\Pi'_{\text{c}}(s^{\text{c}}_{t+1}; \Theta'_{\text{c}})$ are a copy of $\Pi_{\text{r}}$ and $\Pi_{\text{c}}$, with their own parameters $\Theta'_{\text{r}}$ and $\Theta'_{\text{c}}$, respectively, which are used to generate the next actions $\boldsymbol{a}^{\text{r}}_{t+1}$ and $a^{\text{c}}_{t+1}$. Similarly, the target network $Q'(s^{\text{q}}_{t+1}, \boldsymbol{a}^{\text{q}}_{t+1}; \Theta'_{\text{q}})$ of the critic $Q$ is a copy of $Q$ with parameters $\Theta'_{\text{q}}$, which is used for generating the target cumulative return when training $Q$.

\begin{algorithm}[t]
\SetKwInOut{Input}{input}
\SetKwInOut{Output}{output}
\caption{Training of TIARec}
\label{alg:training}

\Input {The training dataset $\{ \mathcal{V}^{u}\}$, learning rate $\lambda$, fusion coefficient $\tau \in [0,1]$ } 
\Output {The parameters $\Theta_{\text{r}}$, $\Theta_{\text{c}}$, and $\Theta_{\text{q}}$ }
		
{Randomly initialize $\Theta_{\text{r}}$, $\Theta_{\text{c}}$, and $\Theta_{\text{q}}$}\;
{Initialize the target networks: $\Theta'_{\text{r}} = \Theta_{\text{r}}$, $\Theta'_{\text{c}} = \Theta_{\text{c}}$, $\Theta'_{\text{q}} = \Theta_{\text{q}}$}\;

{Initialize the capacity of replay buffer $\mathcal{B}$}\;

\While{not convergent}{
	\ForEach {$\mathcal{V}^{u}$}{
		Initialize $\mathcal{O}_0 = \mathcal{M}_0 = \mathcal{N}_0 = \{\}$;
		
        		\For{$t \in \{0,1,\cdots,T\}$}{
			Generate the prototype item $\boldsymbol{a}^{\text{p}}_t$ and the recommended item $\boldsymbol{a}^{\text{r}}_t$ using Equations (\ref{eq:prototype}) and (\ref{eq:r_action}), respectively\;
			Set recommender reward $R^{\text{r}}_t$ using Equation (\ref{eq:r_reward})\;
		        Update $\mathcal{O}_{t+1} = \mathcal{O}_t \cup \{ \boldsymbol{a}^{\text{r}}_t \}$\;
			Generate classifier action $a^{\text{c}}_t$ using Equation (\ref{eq:c_action})\;
			Update $\mathcal{N}_{t+1} = \mathcal{N}_t \cup \{ \boldsymbol{a}^{\text{r}}_t \}$, $\mathcal{M}_{t+1} = \mathcal{M}_t $ with probability $a^{\text{c}}_t$; or $\mathcal{N}_{t+1} = \mathcal{N}_t $, $\mathcal{M}_{t+1} = \mathcal{M}_t \cup \{ \boldsymbol{a}^{\text{r}}_t \}$ with probability $1-a^{\text{c}}_t$\;
			Set classifier reward $R^{\text{c}}_t$ using Equation (\ref{eq:c_reward})\;
			Set overall reward $r_t$ using Equation (\ref{eq:reward})\;
			Set $s_t = \{  \mathcal{O}_t, \mathcal{N}_t, \mathcal{M}_t\}$ and $s_{t+1} = \{  \mathcal{O}_{t+1}, \mathcal{N}_{t+1}, \mathcal{M}_{t+1}\}$\;
			Save transition sample $(s_t, \boldsymbol{a}^{\text{r}}_t, r_t, s_{t+1})$ to $\mathcal{B}$\;
			
			Sample a minibatch from $\mathcal{B}$\;
			Compute the losses $L_{\text{q}}$, $L_{\text{r,c}}$ based on the minibatch, using Equations (\ref{eq:Lq}) and (\ref{eq:Lrc}), respectively\;
			Update the critic $\Theta_{\text{q}} = \Theta_{\text{q}} - \lambda \nabla_{\Theta_{\text{q}}}L_{\text{q}}$\;
			Update the recommender $\Theta_{\text{r}} = \Theta_{\text{r}} - \lambda \nabla_{\Theta_{\text{r}}}L_{\text{r,c}}$\;
			Update the classifier $\Theta_{\text{c}} = \Theta_{\text{c}} - \lambda \nabla_{\Theta_{\text{q}}}L_{\text{r,c}}$\;

            		Update the target networks:
			$\Theta'_{\text{q}} =  \tau \Theta_{\text{q}} + (1-\tau)\Theta'_{\text{q}} $;
			$\Theta'_{\text{r}} =  \tau \Theta_{\text{r}} + (1-\tau)\Theta'_{\text{r}} $;
			$\Theta'_{\text{c}} =  \tau \Theta_{\text{c}} + (1-\tau)\Theta'_{\text{c}} $\;
            	}
        		
	}
	
}

\end{algorithm}

\subsubsection{Loss Function of The Critic}
We expect the estimated cumulative return $\widehat{q}_t$ provided by the critic $Q$ should be as close to the target cumulative return $q_t$ as possible. Let $r_t$ be the total immediate reward and $\{s^{\text{q}}_t, \boldsymbol{a}^{\text{q}}_t, r_t, s^{\text{q}}_{t+1} \}$ be a training sample, then
the loss function of the critic network can be defined as 
\begin{equation}
L_{\text{q}} (\Theta_{\text{q}}) = \mathbb{E}_{s_t, \boldsymbol{a}^{\text{q}}_t, r_t, s_{t+1}} [ q_t - Q (s^{\text{q}}_t, \boldsymbol{a}^{\text{q}}_t = \boldsymbol{a}^{\text{r}}_t \oplus a^{\text{c}}_t; \Theta_{\text{q}} ) ].
\label{eq:Lq}
\end{equation} 
The total immediate reward $r_t$ is defined as
\begin{equation}
r_t = R^{\text{r}}_t + \alpha R^{\text{c}}_t,
\label{eq:reward}
\end{equation}
where $R^{\text{r}}_t$ and $R^{\text{c}}_t$ are the immediate rewards of the recommender agent and the classifier agent, respectively, and $\alpha$ is a hyper-parameter for balancing the contributions of the immediate rewards. 

The target cumulative return $q_t$ is defined as the sum of the immediate reward and the cumulative return of $\{s^{\text{q}}_{t+1}, \boldsymbol{a}^{\text{q}}_{t+1}\}$, 
\begin{equation}
q_t = r_t + \gamma Q'(s^{\text{q}}_{t+1}, \boldsymbol{a}^{\text{q}}_{t+1}; \Theta'_{\text{q}}),
\end{equation}
where $ \gamma \in [0,1]$ is the discount factor. The next action $\boldsymbol{a}^{\text{q}}_{t+1}$ $=$ $ \boldsymbol{a}^{\text{r}}_{t+1} \oplus$$a^{\text{c}}_{t+1}$, where $\boldsymbol{a}^{\text{r}}_{t+1}$ $=$ $\Pi'_{\text{r}}(s^{\text{r}}_{t+1};$ $\Theta'_{\text{r}})$ and $a^{\text{c}}_{t+1}$ $=$ $\Pi'_{\text{c}}(s^{\text{c}}_{t+1}; \Theta'_{\text{c}})$.
As we will see later, in each iteration of the training, the parameters $\Theta'_{\text{q}}$, $\Theta'_{\text{r}}$, and $\Theta'_{\text{c}}$ of the target networks are fixed with the model parameters $\Theta_{\text{q}}$, $\Theta_{\text{r}}$, and $\Theta_{\text{c}}$ of previous iteration.

\subsubsection{Loss Function of The Agents}
As we have seen before, the recommender agent actually generates a prototype item embedding $\boldsymbol{a}^{\text{p}}_t$, while a real item embedding $\boldsymbol{a}^{\text{r}}_t$ is fed to the critic. To connect the recommender agent and the critic for joint training, we need to make them as similar as possible, and at the same time, both agents seek to maximize the expected cumulative return $\widehat{q}_t$. These ideas lead to the following loss function of the agents:
\begin{equation}
L_{\text{r,c}} (\Theta_{\text{r}}, \Theta_{\text{c}}) =  \mathbb{E}_{s_t} \Big[ (\boldsymbol{a}^{\text{r}}_t - \boldsymbol{a}^{\text{p}}_t)^2 - Q \big(s^{\text{q}}_t, \boldsymbol{a}^{\text{q}}_t = \boldsymbol{a}^{\text{p}}_{t} \oplus a^{\text{c}}_{t} \big) \Big].
\label{eq:Lrc}
\end{equation}
Note that since the first term ensures the recommended item embedding $\boldsymbol{a}^{\text{r}}_t$ will be close to the prototype embedding $\boldsymbol{a}^{\text{p}}_t$, we use the prototype item $\boldsymbol{a}^{\text{p}}_{t}$ instead of $\boldsymbol{a}^{\text{r}}_{t}$ to compute the expected cumulative return, which is slightly different from Equation (\ref{eq:Lq}) where we use $\boldsymbol{a}^{\text{r}}_{t}$.

\subsubsection{Training Algorithm}

The training of TIARec uses an offline strategy shown in Algorithm \ref{alg:training}, where each iteration is composed of two stages, the sample generating and the policy updating. In the sample generating stage of each time step $t$ (lines 8-16), the recommender agent first recommends an item $\boldsymbol{a}^{\text{r}}_t$ (line 8), and the classifier agent generates a probability $a^{\text{c}}_t$ (line 11) with respect to which the interaction with the recommended item is classified as noise or normal (line 12). Then we set the overall immediate reward $r_t$ (line 14) by fusing the immediate rewards $R^{\text{r}}_t$ (line 9) and $R^{\text{c}}_t$ (line 13) of the recommender agent and the classifier agent. At last, the state is updated (line 15) and the transition sample is stored to the replay buffer (line 16). In the policy updating stage (lines 17-22), we first sample a minibatch from the replay buffer (line 17), and then use gradient descent to update the parameters of the critic, the recommender agent, the classifier agent, and the target networks (lines 19-22). 

\begin{table}[!t]
\caption{Statistics of the datasets}
\label{table1}
\centering
\begin{tabular}{c|ccc} \hline
 &Movielens-1M&Beauty&Tianchi\\ \hline
The number of users & 6,040& 1,210,271&6,471\\ 
The number of items & 3,952& 249,274&554,442\\ 
The number of ratings & 1,000,209 & 2,023,096&802,757\\ \hline
\end{tabular}
\end{table}

\section{EXPERIMENTS}

The goals of the experiments are to answer the following three research questions: 
\begin{itemize}
\item \textbf{RQ1} how does TIARec perform as compared to the state-of-the-art baselines?
\item \textbf{RQ2} how does the classifier agent contribute to the performance of TIARec?
\item \textbf{RQ3} How can the superiority of TIARec be illustrated with an intuitive and visualizable case study?
\item \textbf{RQ4} How is the robustness of TIARec?
\item \textbf{RQ5} how do the hyper-parameters affect the performance?
\end{itemize}
The implementation of TIARec is available on https:// github.com/dududu123456/TIARec.

\subsection{Experimental Setup}
\subsubsection{Datasets}
We adopt the following three public datasets for the experiments.
\begin{itemize}
\item \textbf{Movielens-1M\footnote{https://grouplens.org/datasets/movielens/} } Movielens-1M is a popular dataset for evaluating recommendation models, which contains over 1 million ratings given by 6,040 users to 3,952 movies.
\item \textbf{Beauty\footnote{https://www.kaggle.com/skillsmuggler/amazon-ratings} } Beauty contains over 2 million ratings given by more than 1 million users to 249,274 products on Amazon.
\item \textbf{Tianchi\footnote{https://tianchi.aliyun.com/dataset/dataDetail?dataId=64345} } Tianchi contains over 802,757 ratings given by 6,471 users to 554,442 products on TMALL, a well-known Chinese e-commerce website.

\end{itemize}
The statistics of the datasets is given in Table \ref{table1}. On each dataset, we use the first 70\% of the data as the training set, the middle 10\% as the validation set, and the remaining 20\% as the testing set.

\subsubsection{Baselines}
As we have mentioned, TIARec is a RL-based robust recommendation model that can capture users' temporary interest. Therefore, to verify our claimed contributions, we compare TIARec with seven representative methods including two robust recommendation models (CDAE and APR), four RL-based models (DEERs, HRL, KGQR, and PDQ), and one interactive recommendation model (I-CARS), which are briefly introduced as follows:

\begin{itemize}
\item \textbf{CDAE} 
CDAE \cite{CDAE} is a Denoising Auto-encoder based collaborative filtering model, which can learn robust latent representations of corrupted user-item interactions for top-$k$ recommendation.

\item \textbf{APR} 
APR \cite{APR} is an Adversarial Personalized Ranking framework, which offers the recommendation robustness by injecting adversarial noise to the parameters of a pairwise ranking model BPR \cite{Rendle2012BPR} during an adversarial training.

\item \textbf{DEERs} DEERs \cite{DEERs} is a deep reinforcement learning based recommendation model, which can simultaneously model positive and negative feedbacks using a Deep Q-network (DQN) \cite{DQN} with a pairwise training.

\item \textbf{HRL} HRL \cite{zhang2019hierarchical} is a hierarchical reinforcement learning model for recommendations of online courses, which directly remove atypical interactions through a hierarchical sequential decision process.

\item \textbf{KGQR} KGQR \cite{Zhou2020} is a knowledge graph enhanced Q-learning model for interactive recommendations, which incorporates the prior knowledge of the item correlation learned from knowledge graph into the sequential decision making to improve sample efficiency for learning better representations of items and users.

\item \textbf{PDQ} PDQ \cite{zou2020pseudo} is another deep Q-learning based interactive recommendation model, where the recommendation policy is trained without the requirement of real customer interactions, but with a pseudo environment model that simulates the feedbacks of users.

\item \textbf{I-CARS} I-CARS \cite{lumbantoruan2019cars} is an interactive context-aware recommendation model, which can continually update its parameters according to a user's recent preferences that are revealed by the user's feedbacks to the deliberately selected items.

\end{itemize}

Additionally, in order to verify the effectiveness of the classifier agent of TIARec, we also compare TIARec with one more baseline method TIARec-C, which is a variant of TIARec where the classifier agent is removed.

\subsubsection{Evaluation Protocol}
We choose the widely used Hit Rate (HR), Recall, and NDCG to evaluate the performance of TIARec and the baselines. Let $\{ v^u_1, \cdots, v^u_{N_u} \}$ be the testing set consisting of $N_u$ ground-truth items that user $u$ has interacted with. For each testing instance $v^u_i$ ($1 \le i \le N_u$), we generate a ranking list $\boldsymbol{l}^u_i$ of items using TIARec. Then the number of $u$'s cases that a ground-truth is ranked in the top-$k$ items can be computed as
\begin{equation}
S^u_k = \sum_{i=1}^{N_u}{\mathbb{I}\big(rank(v^u_i, \boldsymbol{l}^u_i)} \leq k \big),
\end{equation}
where $rank(v, l)$ is the rank of item $v$ in list $l$, and $\mathbb{I}(x)$ is the indicator function that returns 1 if $x$ is true, otherwise 0. Then the metrics can be defined as follows:
\begin{equation} 
	{\rm HR}@k = \frac{1}{|\mathcal{U}|}{\sum_{u \in \mathcal{U}}{\mathbb{I}( S^u_k\geq 1)}},
\end{equation}
\begin{equation} 
	{\rm Recall}@k = \frac{1}{|\mathcal{U}|}\sum_{u \in \mathcal{U}}{\frac{S^u_k}{N_{u}}},
\end{equation}
\begin{equation}
	{\rm NDCG}@k = \frac{1}{|\mathcal{U}|}\sum_{u \in \mathcal{U}}{\frac{1}{N_{u}} \sum_{i=1}^{N_{u}}{\frac{\mathbb{I}\big(rank(v^u_i, \boldsymbol{l}^u_i)\leq k \big)}{\log_2{\big(1+rank(v^u_i, \boldsymbol{l}^u_i) \big)}}}}.
\end{equation}

\subsubsection{Hyper-parameter Setting}
The hyper-parameters are tuned on the validation sets. The embedding dimensionality $d$ is set to 64 for Movielens-1M and 128 for Beauty and Tianchi. The reward balance factor $\alpha$ in Equation (\ref{eq:reward}) is set to 0.1 for Movielens-1M,  0.5 for Beauty and 1.0 for Tianchi. On three datasets, the discount rate $\gamma$, the size of replay buffer $\mathcal{B}$, and the fusion coefficient $\tau$ are set to 0.99, 2,000 and 0.01, respectively. For fairness, the hyper-parameters of the baselines are also set to their optimal values on the validation sets. %One can refer to the appendix for the details.

\begin{table*}
\caption{Performance comparison on Movielens-1M}
\label{tbl:resultMovielens}
\begin{adjustbox}{center}
\begin{tabular}{l|c|c|c|c|c|c|c|c|c} 
\toprule
\multirow{2}{*}{\diagbox{Method}{Metirc}}&
\multicolumn{3}{c|}{HR@\textit k}&
\multicolumn{3}{c|}{Recall@\textit k}&
\multicolumn{3}{c}{NDCG@\textit k}\\
\cline{2-10}
  &\textit k=5&\textit k=10&\textit k=20&\textit k=5&\textit k=10&\textit k=20 &\textit k=5&\textit k=10&\textit k=20 \\ \hline
 CDAE&0.0554&0.0561&0.3014&0.0134&0.0297&0.0674&0.0003&0.0044&0.0187\\
 APR &0.0536&0.0613 &0.3387& 0.0121&0.0315&0.0752 & 0.0034&0.0151&0.0304\\ \hline
DEERs&0.0815&0.1166 &0.3651& 0.0259	&0.0477&0.1134& 0.0083&0.0174&0.0381\\ 
HRL &0.0734&0.1284 &0.3433 & 0.0299&0.0311&0.0998&0.0087&0.0188 &0.0377 \\ 
KGQR&0.1391&0.2765 &0.5029&0.0283&0.0519  &0.1378  &0.0118 &0.0165&0.0545  \\
PDQ&0.0829&0.1619&0.3712&0.0174&0.0418&0.1049&0.0126&0.0196&0.0364\\ \hline
I-CARS&0.0629&0.0929&0.3406&0.0150&0.0356&0.1009&0.0025&0.0089&0.0219\\ \hline
TIARec-C& 0.1104&0.1916&0.4128& 0.0231&0.0505&0.1043&0.0072&0.0170 &0.0325\\
TIARec&\textbf{0.1706}& \textbf{0.3110}& \textbf{0.5793}&\textbf{0.0322}& \textbf{0.0583}&\textbf{0.1602}&\textbf{0.0151}&\textbf{0.0274}&\textbf{0.0701}\\  
Improv. &22.65\%&12.48\%&15.19\%&7.69\%&12.33\%&12.26\%&19.84\%&39.80\%&28.62\% \\
\bottomrule
\end{tabular}
\end{adjustbox}
\end{table*}

\begin{table*}
\caption{Performance comparison on Beauty}
\label{tbl:resultBeauty}
\begin{adjustbox}{center}
\begin{tabular}{l|c|c|c|c|c|c|c|c|c} 
\toprule
\multirow{2}{*}{\diagbox{Method}{Metirc}}&
\multicolumn{3}{c|}{HR@\textit k}&
\multicolumn{3}{c|}{Recall@\textit k}&
\multicolumn{3}{c}{NDCG@\textit k} \\
\cline{2-10}
  &\textit k=5&\textit k=10&\textit k=20&\textit k=5&\textit k=10&\textit k=20&\textit k=5&\textit k=10&\textit k=20\\ \hline
CDAE &0.0024&0.0936&0.2431&0.0009&0.0161&0.0339&0.0017&0.0054&0.0151\\
APR &0.0085&0.1077&0.3668&0.0018&0.0137&0.0548&0.0018&0.0081&0.0216\\ \hline
DEERs&0.1025& 0.1902&0.3066& 0.0054&0.0217&0.0514& 0.0041&0.0077&0.0184\\
HRL &0.1033&0.1178&0.2433 &0.0105&0.0233 &0.0531 & 0.0049&0.0118 &0.0197\\ 
KGQR&0.1120&0.1714& 0.3731& 0.0114&0.0183 & 0.0425 &0.0056&0.0152 & 0.0166 \\
PDQ&0.1328&0.2074&0.4378&0.0160&0.0212&0.0548&0.0073&0.0124&0.0183\\\hline
I-CARS&0.1044&0.1704&0.3728&0.0104&0.0177&0.0357&0.0021&0.0071&0.0163\\\hline
TIARec-C&0.1009&0.2027& 0.3101& 0.0151&0.0229&0.0409& 0.0071&0.0130&0.0173\\
TIARec&\textbf{0.1652}&\textbf{0.2439}& \textbf{0.4699}&\textbf{0.0231}& \textbf{0.0328}&\textbf{0.0701}&\textbf{0.0104}&\textbf{0.0207}&\textbf{0.0304} \\ 
Improv. &24.40\%&17.60\%&7.33\%&44.38\%&40.77\%&27.92\%&42.47\%&36.18\%&40.74\% \\
\bottomrule
\end{tabular}
\end{adjustbox}
\end{table*}

\begin{table*}
\caption{Performance comparison on Tianchi}
\label{tbl:resultTianchi}
\begin{adjustbox}{center}
\begin{tabular}{l|c|c|c|c|c|c|c|c|c} 
\toprule
\multirow{2}{*}{\diagbox{Method}{Metirc}}&
\multicolumn{3}{c|}{HR@\textit k}&
\multicolumn{3}{c|}{Recall@\textit k}&
\multicolumn{3}{c}{NDCG@\textit k}\\
\cline{2-10}
  &\textit k=5&\textit k=10&\textit k=20&\textit k=5&\textit k=10&\textit k=20&\textit k=5&\textit k=10&\textit k=20\\ \hline
 CDAE&0.0034&0.0172 &0.1669 & 0.0011&0.0073&0.0218&0.0003&0.0017 &0.0083\\
 APR &0.0079& 0.0745& 0.1967&0.0028&0.0109&0.0332&0.0007&0.0053 &0.0088\\ \hline
DEERs&0.0363&0.0818&0.2446& 0.0046&0.0102&0.0366& 0.0014&0.0037&0.0114\\ 

HRL &0.0847&0.0940 & 0.2327& 0.0092&0.0107&0.0344&0.0039&0.0090 &0.0117 \\ 

KGQR&0.0863&0.1417 &0.2632&0.0105&0.0165  &0.0405  &0.0034 &0.0069&0.0131  \\
PDQ&0.0771&0.1255&0.2517&0.0085&0.0120&0.0378&0.0027&0.0036&0.0121\\\hline

I-CARS&0.0238&0.0515&0.1675&0.0006&0.0060&0.0122&0.0003&0.0029&0.0091\\\hline

TIARec-C& 0.0526&0.1037&0.2199& 0.0082&0.0160&0.0392&0.0022&0.0074 &0.0112\\
TIARec&\textbf{0.0922}& \textbf{0.1515}& \textbf{0.2713}&\textbf{0.0119}& \textbf{0.0194}&\textbf{0.0433}&\textbf{0.0062}&\textbf{0.0104}&\textbf{0.0154}\\  
Improv. &8.85\%&6.92\%&3.08\%&13.33\%&17.58\%&6.91\%&58.97\%&15.56\%&17.56\% \\
\bottomrule
\end{tabular}
\end{adjustbox}
\end{table*}

\begin{figure}[!t]
\centering
\includegraphics[scale=0.5]{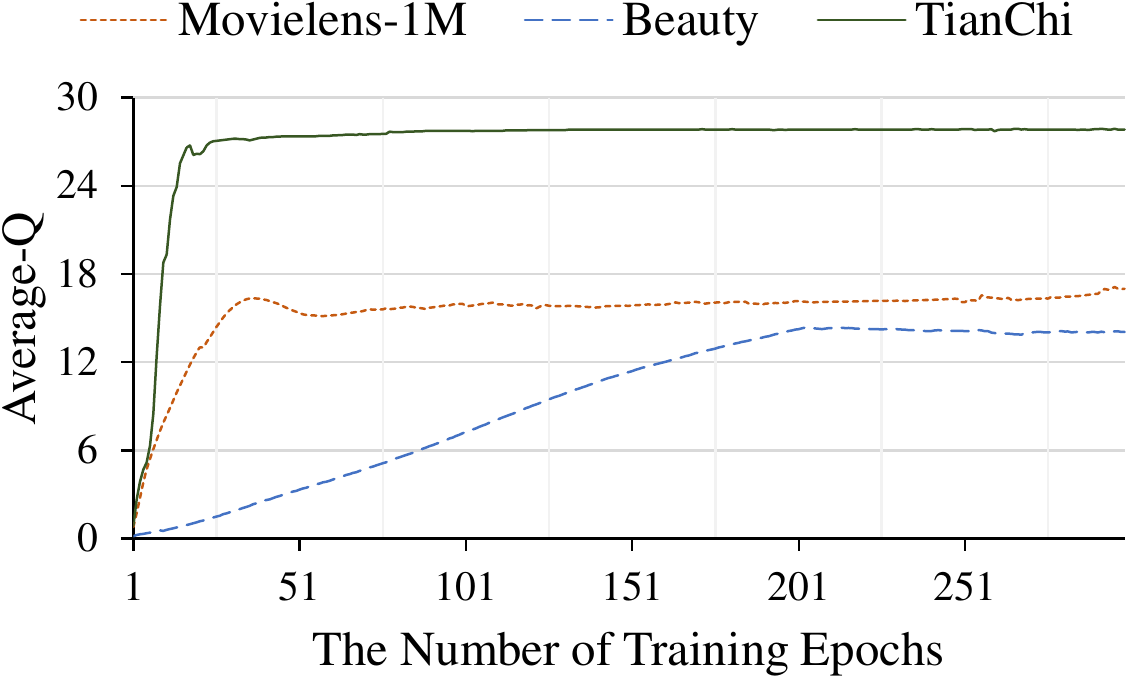}
\caption{Training procedure on all datasets.}
\label{fig:reward}
\end{figure}

\begin{figure*}[!t]
\centering
\subfigure[]{
\begin{minipage}[t]{0.3\textwidth}
\centering
\includegraphics[scale=0.45]{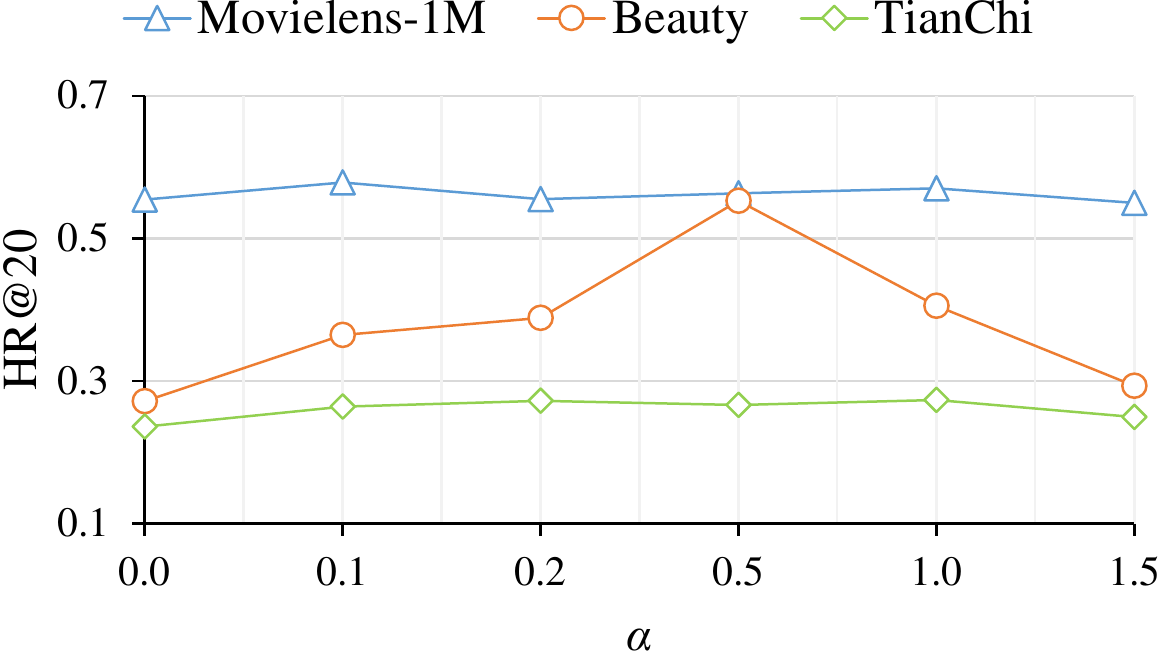}
\end{minipage}%
}
\subfigure[]{
\begin{minipage}[t]{0.3\textwidth}
\centering
\includegraphics[scale=0.45]{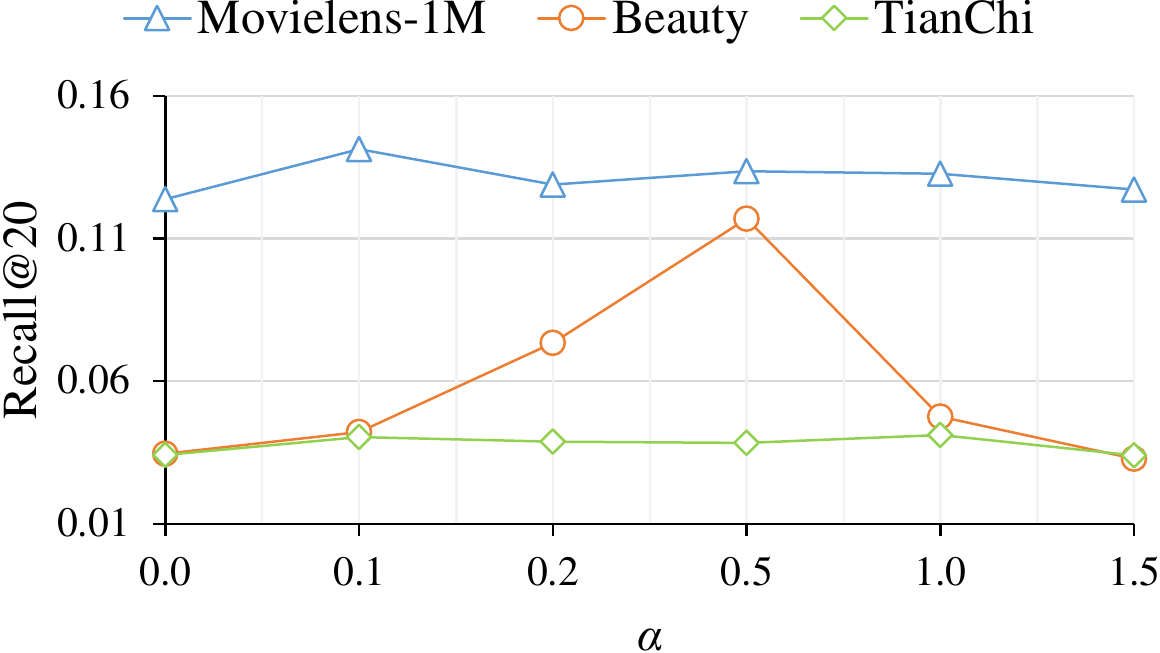}
\end{minipage}
}
\subfigure[]{
\begin{minipage}[t]{0.3\textwidth}
\centering
\includegraphics[scale=0.45]{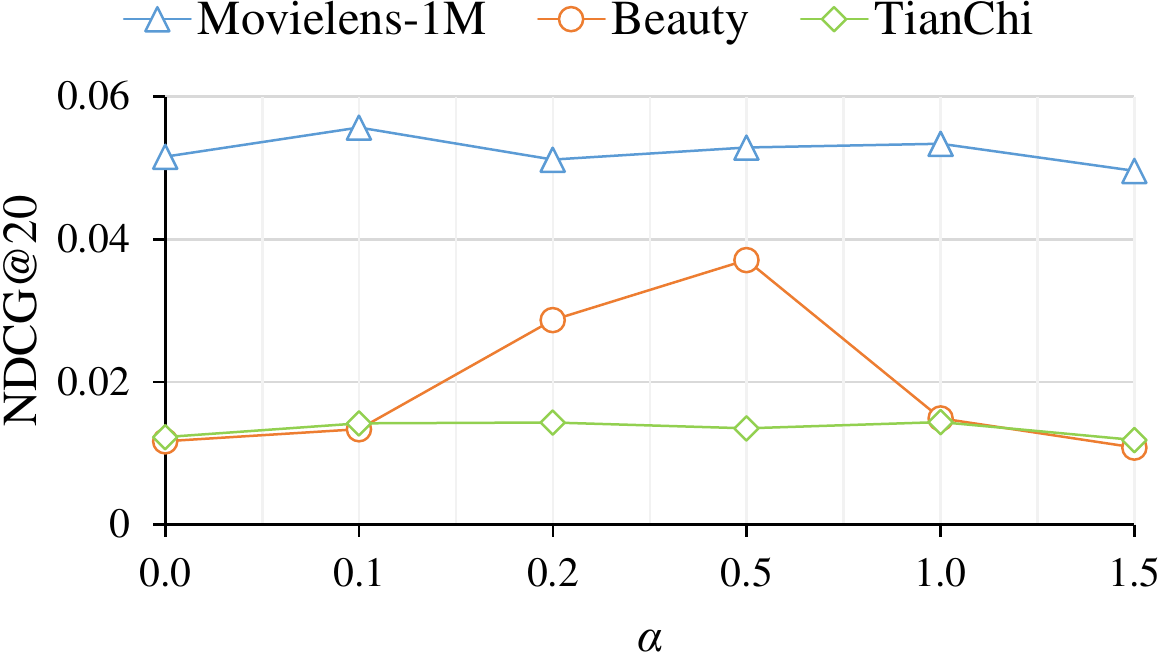}
\end{minipage}
}
\caption{The effectiveness of the classifier agent.}
\label{fig:figure-a}
\end{figure*}

\subsection{Comparative Study (RQ1)}
Tables \ref{tbl:resultMovielens}, \ref{tbl:resultBeauty}, and \ref{tbl:resultTianchi} show the results of the comparison between TIARec and the baselines on Movielens-1M, Beauty, and Tianchi, respectively, where the last row shows the percentage by which TIARec improves the performance of the best baseline. 

At first, we can note that on both datasets TIARec outperforms the robust recommendation models. In contrast to the robust models CDAE and APR, which rely on some kind of noise filtering technique, TIARec can learn from the atypical interactions rather than filtering them. Particularly, TIARec can learn the knowledge about how to capture the users' temporary interest from the atypical interactions, and transfer it to the recommender agent from the classifier agent during the joint training, which makes TIARec more powerful than the traditional robust recommendation models. 

At the same time, we can observe that TIARec achieves significant superiority over the RL-based baselines, which demonstrates the advantages of the classifier agent of TIARec. In particular, unlike the traditional RL-based models, which offers no robustness since they indiscriminately treat every user-item interaction as normal, TIARec can distinguish atypical interactions from normal ones with the help of the classifier agent during the joint training with the recommender agent, which brings two advantages. First, the differential treatment reduces the impact of the atypical interactions for learning users' general preference. Second, the joint training implicitly teaches the recommender agent how to capture users' temporary interest. Such two advantages enable the recommender agent of TIARec to make recommendations with a balance between users' general preference and temporary interest, which contrasts sharply with the traditional RL-based models.

We can also see that TIARec performs better than I-CARS on each dataset. I-CARS aims at capturing a user's preference shift according to the user's feedbacks to the well-chosen probing questions, which is similar to the target of TIARec. However, there are two points that tell I-CARS apart from TIARec. First, I-CARS considers each feedback to be a signal of a user's true preference, which makes I-CARS more susceptible to the noise incurred by the user's feedbacks to the probing questions which cannot reflect the user's true preference. In contrast, the recommender agent of TIARec can learn to discern whether an atypical behavior represents the user's true temporary interest or just a noise, by the joint training with the auxiliary classifier agent. Second, essentially I-CARS is a greedy algorithm which wants to fit every observation, while TIARec pursues the maximization of the long-term return due to the nature of reinforcement learning. For the same reason, almost in each case all the RL-based models achieve a better performance than I-CARS. However, we can see that I-CARS still outperforms CDAE and APR in most cases, due to its sensitivity to user's preference changes.

At last, Figure \ref{fig:reward} illustrates the training process of TIARec in terms of the average cumulative return estimated by the critic over all the users, which is defined as:
\begin{equation}
\text{Average-Q}=\frac{1}{|\mathcal{U}|}\sum_{u \in \mathcal{U}}\frac{1}{T}{\sum_{t=0}^{T}\widehat{q}_{t}^{u} },
\label{eq:Average-Q}
\end{equation}
where $\widehat{q}_{t}^{u} $ is the expected cumulative return estimated by the critic for user $u$ at time step $t$ (see Equation (\ref{eq:critic})). From Figure \ref{fig:reward} we can see that as the number of training epochs increases, the Average-Q first improves, and then approaches convergence after 50 epochs on Movielens-1M, 200 epochs on Beauty, and 30 epochs on Tianchi. Note that it is reasonable that the training on Beauty takes longer time before convergence since it has much greater scale of data.

\begin{figure*}[!t]
\centering
\subfigure[Preference of the sample user]{
\begin{minipage}[t]{0.3\textwidth}
\centering
\includegraphics[scale=0.2]{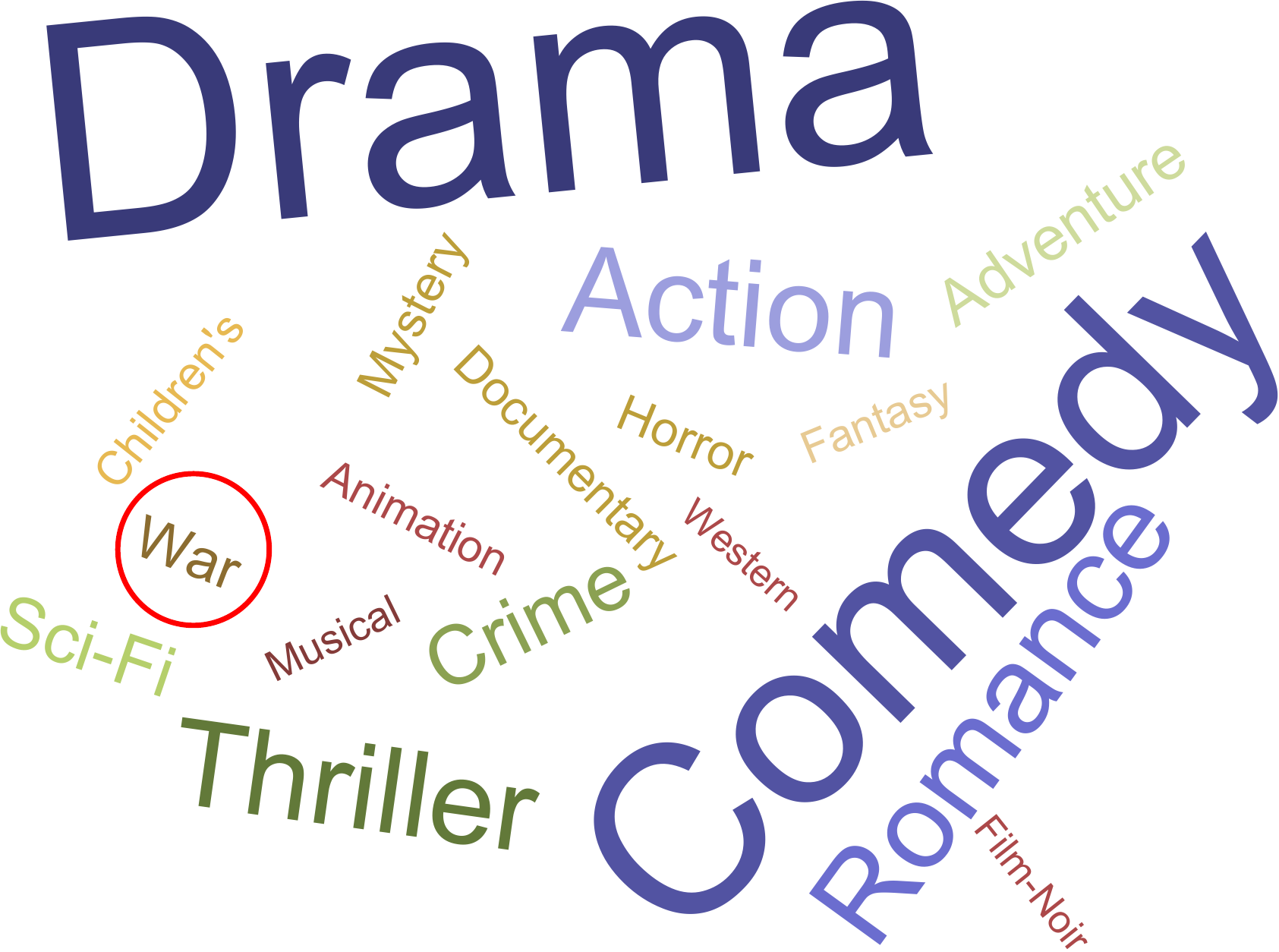}
\label{fig:preference}
\end{minipage}%
}
\subfigure[TIARec]{
\begin{minipage}[t]{0.3\textwidth}
\centering
\includegraphics[scale=0.2]{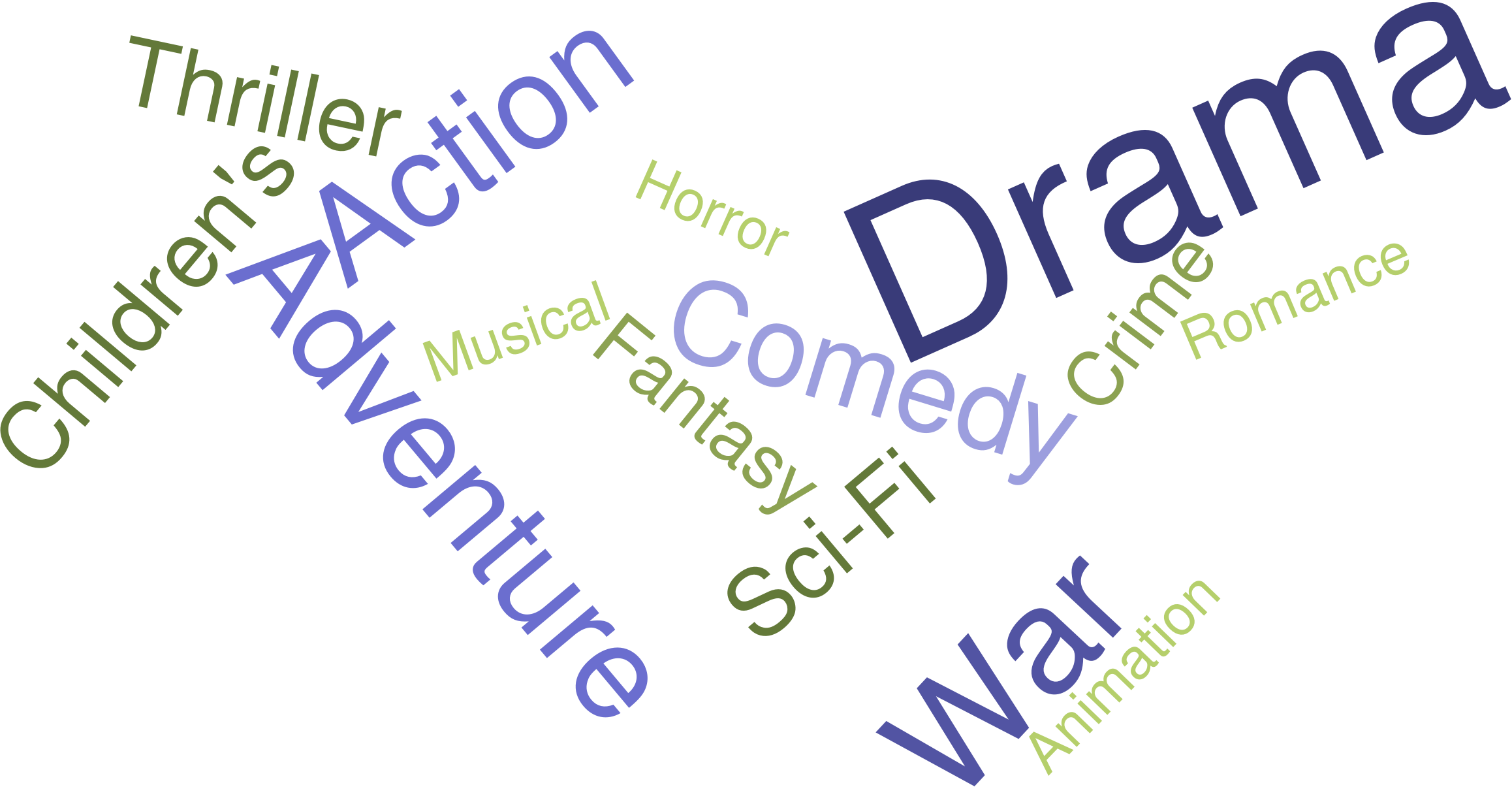}
\label{fig:TIARec}
\end{minipage}
}
\subfigure[TIARec-C]{
\begin{minipage}[t]{0.3\textwidth}
\centering
\includegraphics[scale=0.2]{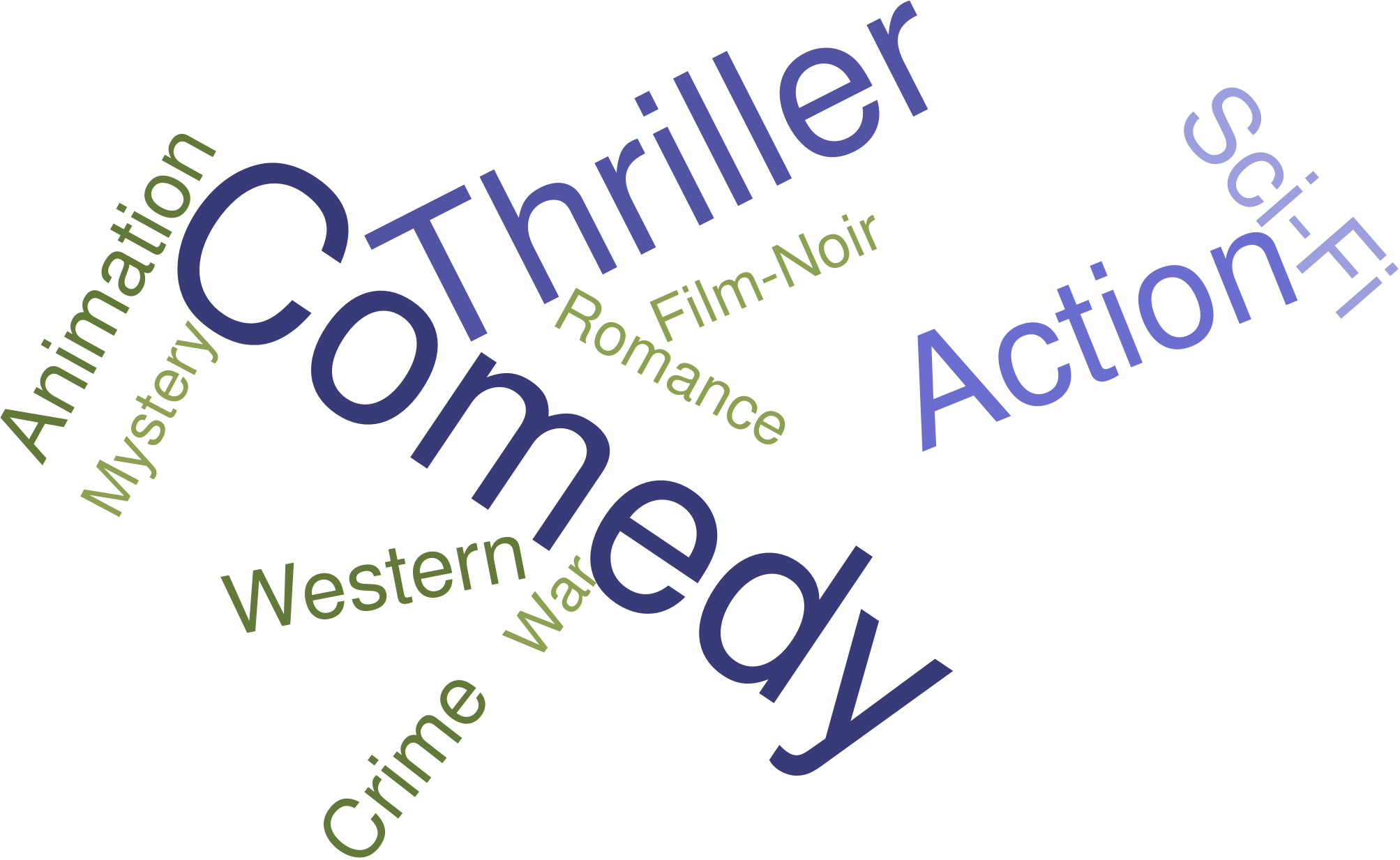}
\label{fig:TIARec-C}
\end{minipage}
}

\subfigure[PDQ]{
\begin{minipage}[t]{0.3\textwidth}
\centering
\includegraphics[scale=0.2]{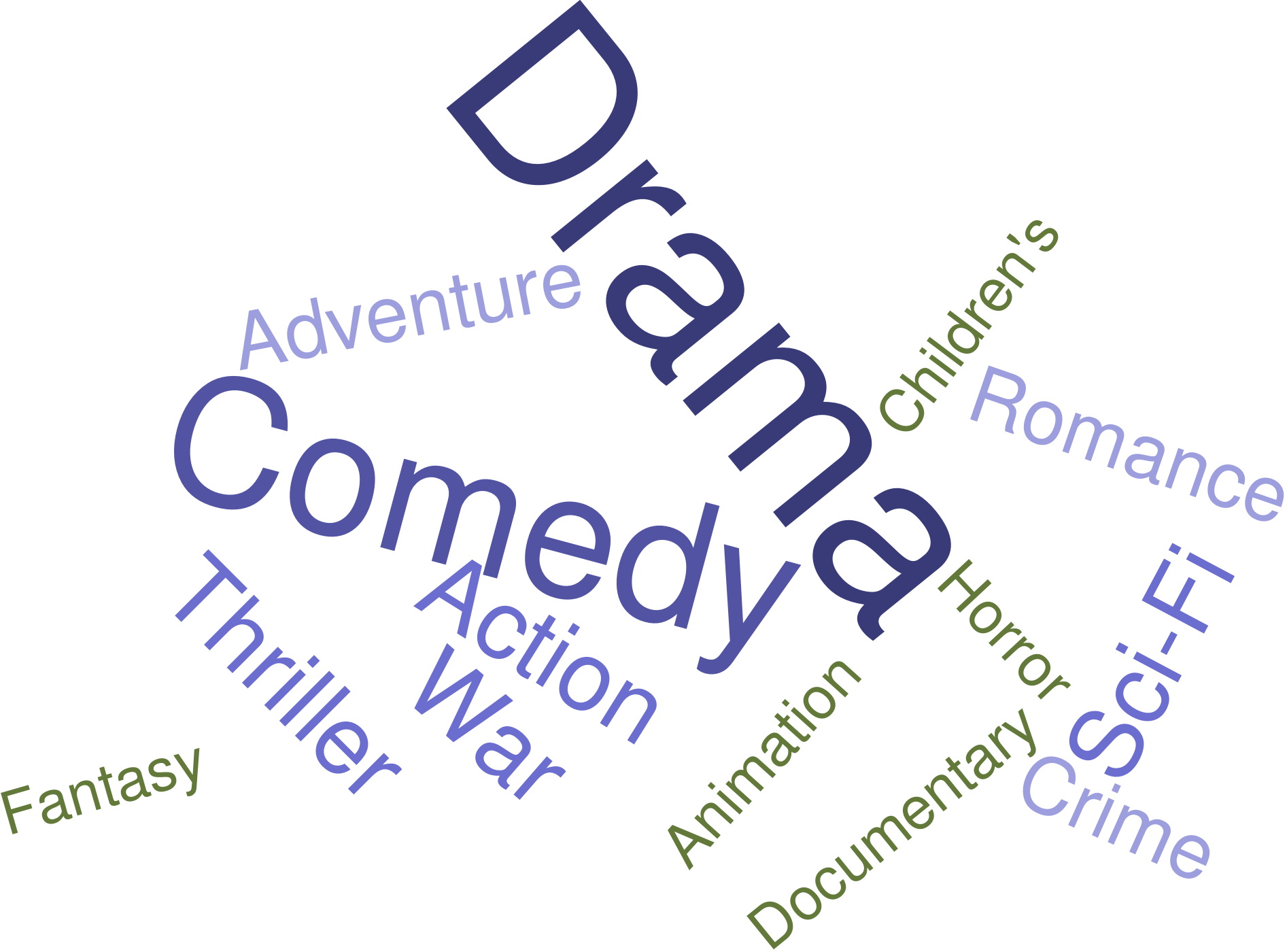}
\label{fig:PDQ}
\end{minipage}
}
\subfigure[KGQR]{
\begin{minipage}[t]{0.3\textwidth}
\centering
\includegraphics[scale=0.2]{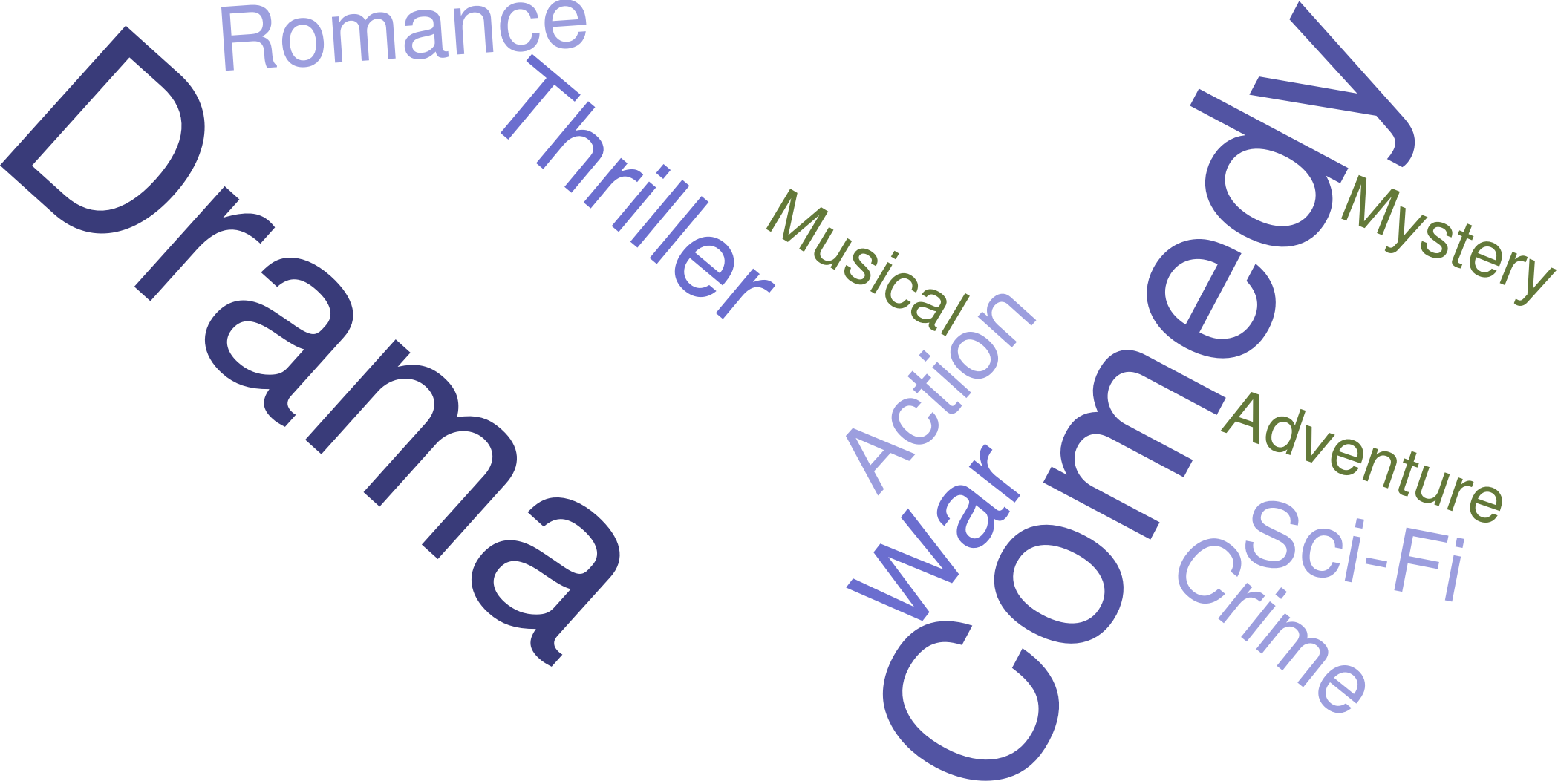}
\label{fig:KGQR}
\end{minipage}
}
\subfigure[DEERs]{
\begin{minipage}[t]{0.3\textwidth}
\centering
\includegraphics[scale=0.2]{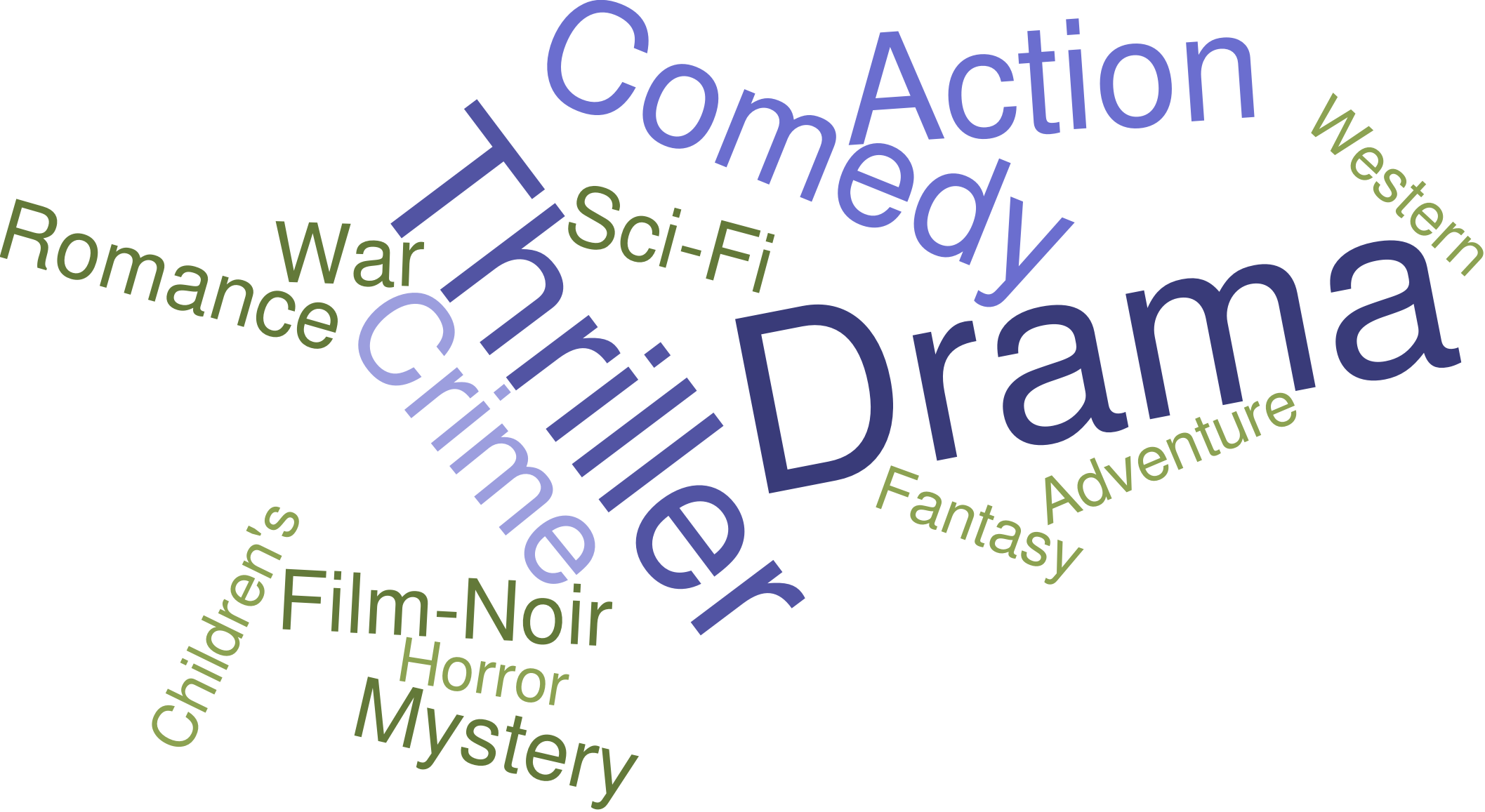}
\label{fig:DEERs}
\end{minipage}
}

\subfigure[HRL]{
\begin{minipage}[t]{0.3\textwidth}
\centering
\includegraphics[scale=0.2]{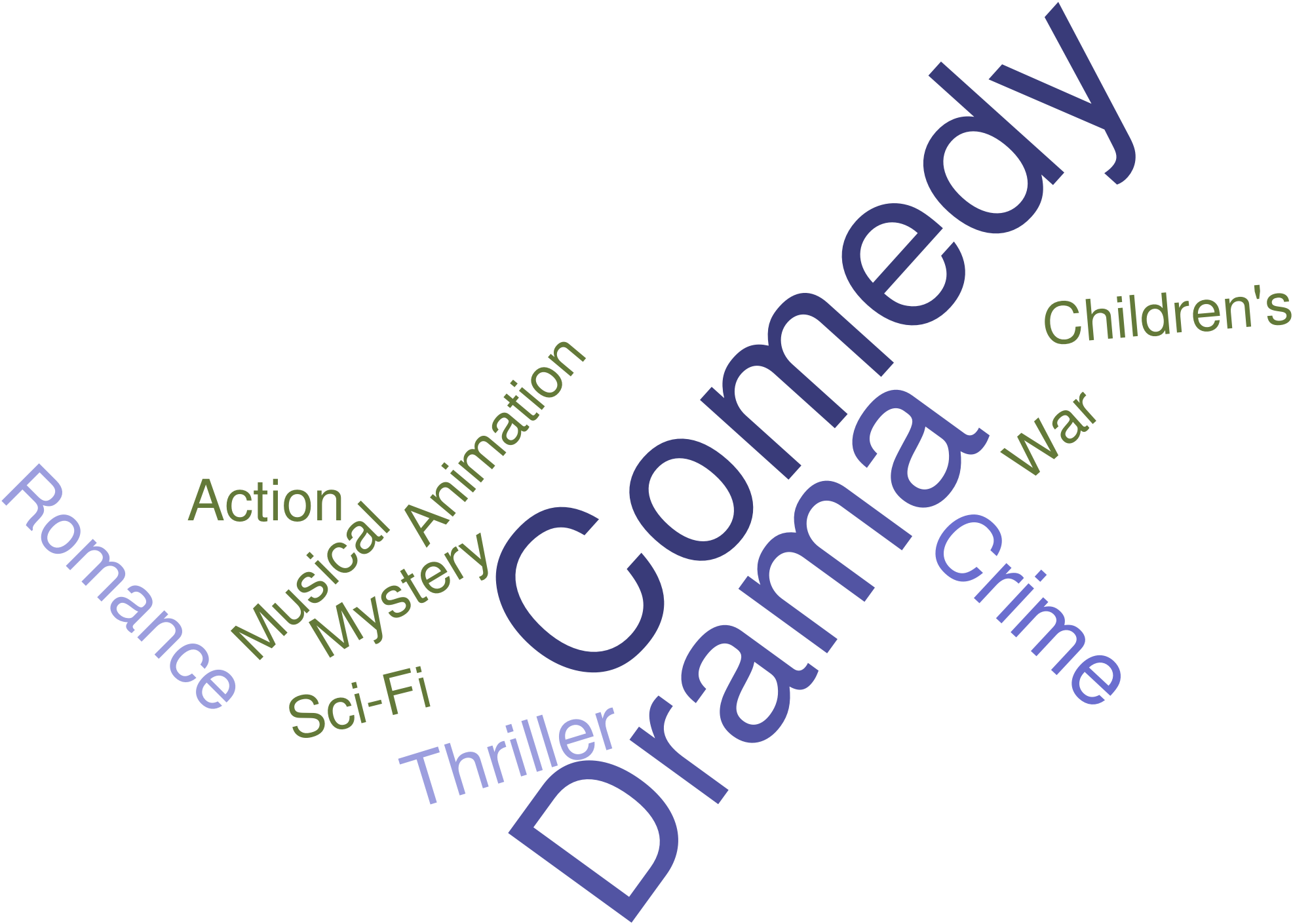}
\label{fig:HRL}
\end{minipage}
}
\subfigure[CDAE]{
\begin{minipage}[t]{0.3\textwidth}
\centering
\includegraphics[scale=0.2]{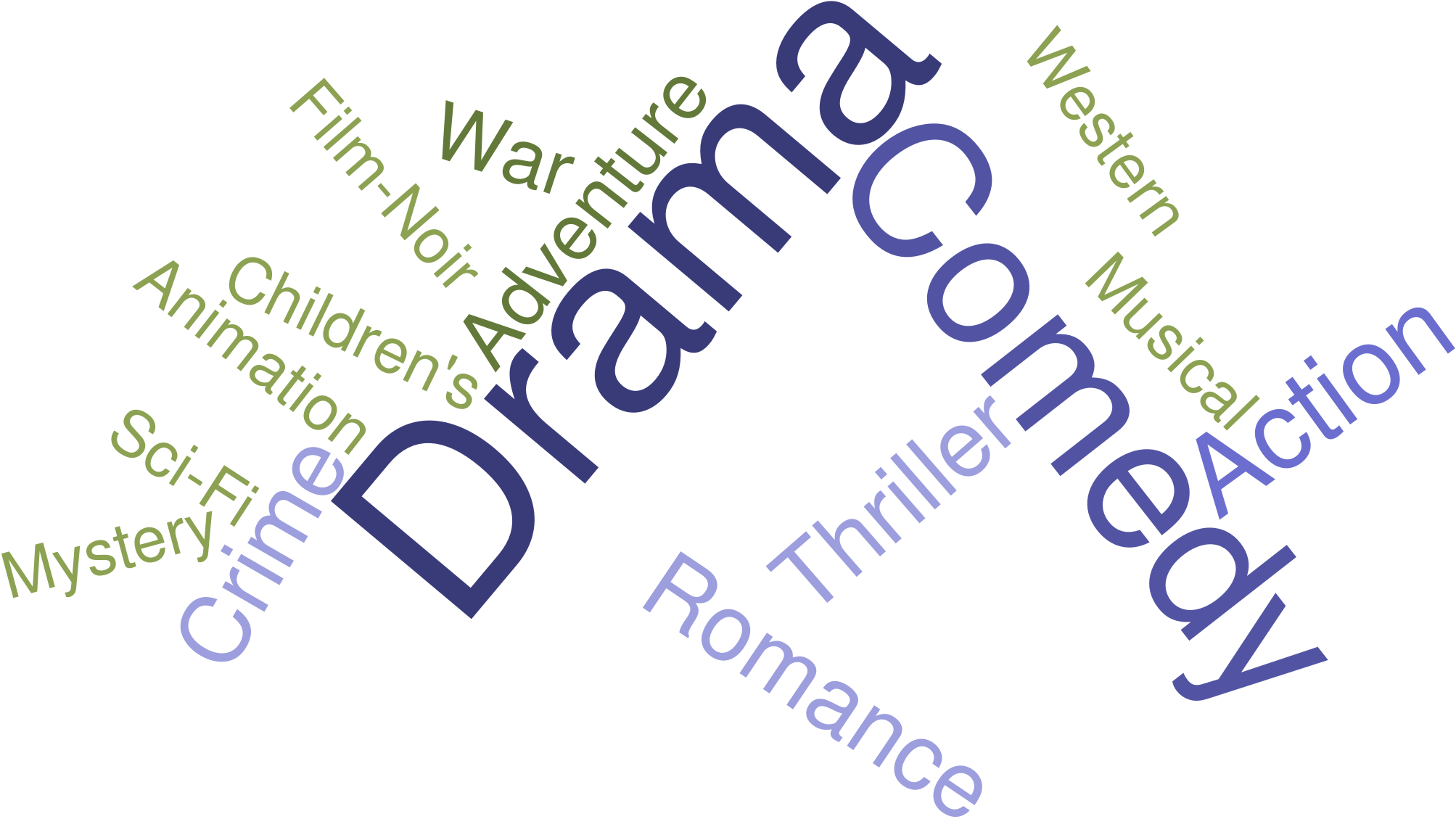}
\label{fig:CDAE}
\end{minipage}
}
\subfigure[APR]{
\begin{minipage}[t]{0.3\textwidth}
\centering
\includegraphics[scale=0.2]{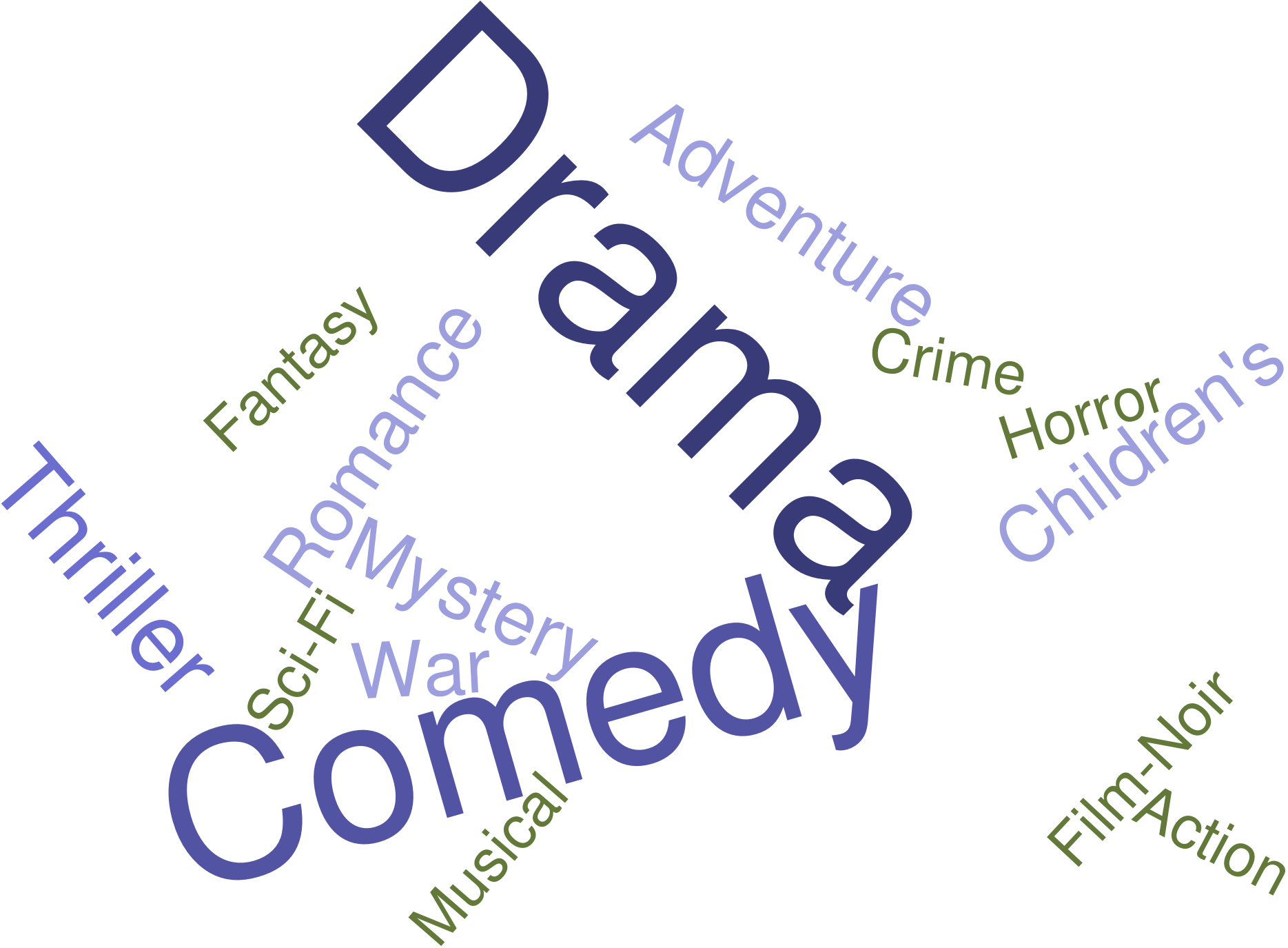}
\label{fig:APR}
\end{minipage}
}

\caption{Case study.}
\label{fig:CaseStudy}
\end{figure*}

\subsection{Ablation Study (RQ2)}

Now we investigate the effectiveness of the classifier agent of TIARec. At first we compare TIARec with its variant TIARec-C where the classifier agent is removed. From Tables \ref{tbl:resultMovielens}, \ref{tbl:resultBeauty}, and \ref{tbl:resultTianchi} we can observe that TIARec considerably improves the performance of the variant TIARec-C on three datasets, which demonstrates the indispensable role of the classifier agent for the robustness of TIARec. As we have mentioned before, the classifier agent can identify whether an interaction is atypical during the joint training with the recommender agent with the objective of maximizing the expected cumulative return estimated by the critic. In fact, once the classifier agent is removed, TIARec degenerates to a traditional model which loses the capability of learning users' temporary interest from atypical interactions. 

The classifier agent plays its role during the training process through its immediate reward in Equation (\ref{eq:reward}), where its weight is regularized by the coefficient $\alpha$. To get further insight into the classifier agent, Figure \ref{fig:figure-a} illustrates how the performance of the recommender agent changes with the different weights $\alpha$ of the classifier agent on both datasets. At first, when $\alpha = 0$, the classifier agent contributes nothing and TIARec degenerates to TIARec-C. As $\alpha$ increases, the classifier agent accounts for more proportion of the total immediate reward and more knowledge about learning temporary interest from atypical interactions is transferred to the recommender agent, which leads to the better performance of the latter. However, when $\alpha$ is greater than 0.1 on Movielens-1M, 0.5 on Beauty, and 1.0 on Tianchi, the curves begin dropping. This is because an overlage $\alpha$ results in the total immediate reward overwhelmed by the classifier agent and consequently causes the recommender agent to pays excessive attention to the users' temporary interest. 

\begin{figure*}[t]
\centering
\subfigure[Movielens-1M]{
\begin{minipage}[t]{0.307\textwidth}
\centering
\includegraphics[scale=0.42]{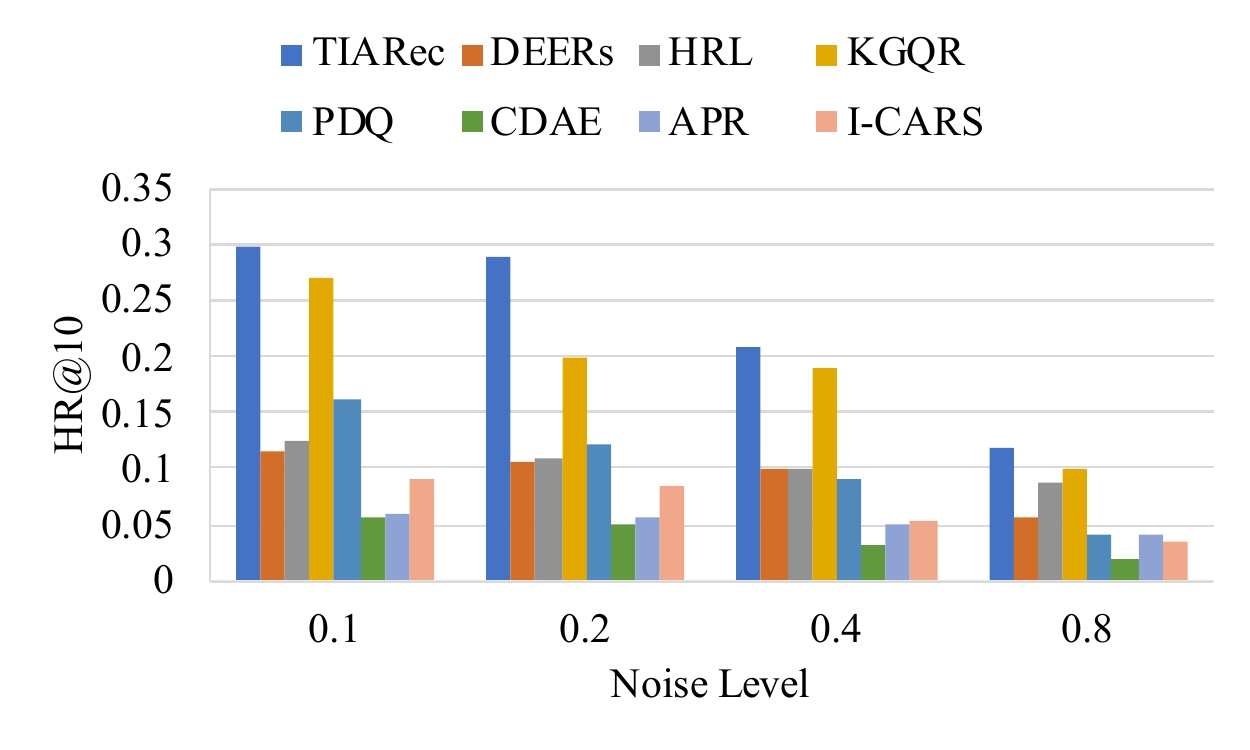}
\end{minipage}%
}
\subfigure[Beauty]{
\begin{minipage}[t]{0.3\textwidth}
\centering
\includegraphics[scale=0.42]{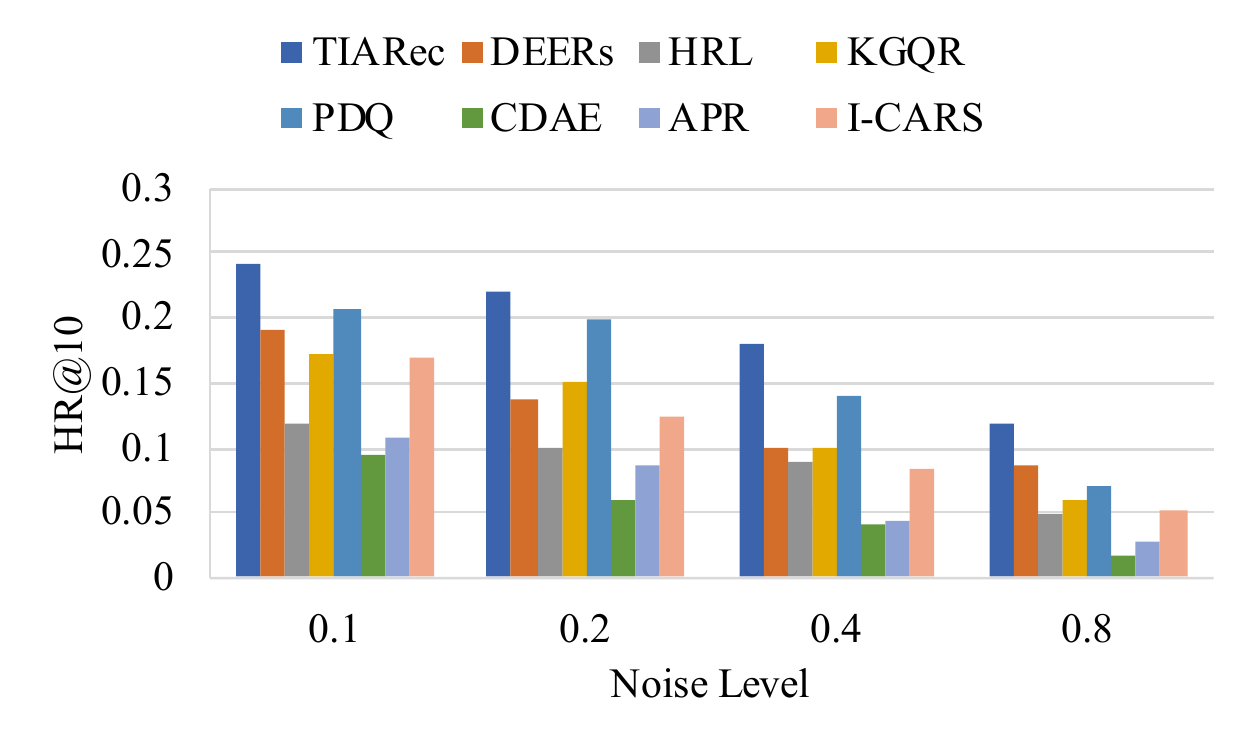}
\end{minipage}
}
\subfigure[Tianchi]{
\begin{minipage}[t]{0.3\textwidth}
\centering
\includegraphics[scale=0.42]{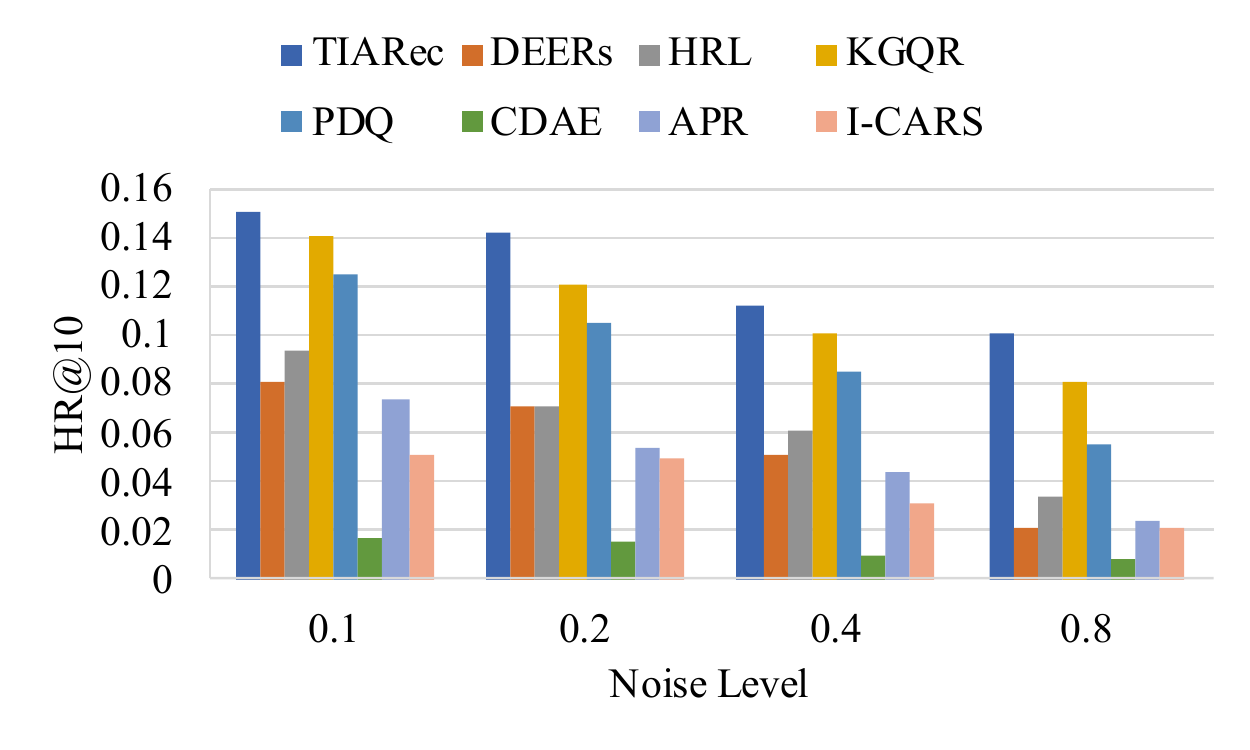}
\end{minipage}
}
\caption{Robustness over different noise levels.}
\label{fig:robust}
\end{figure*}

\begin{figure*}[t]
\centering
\subfigure[]{
\begin{minipage}[t]{0.3\textwidth}
\centering
\includegraphics[scale=0.45]{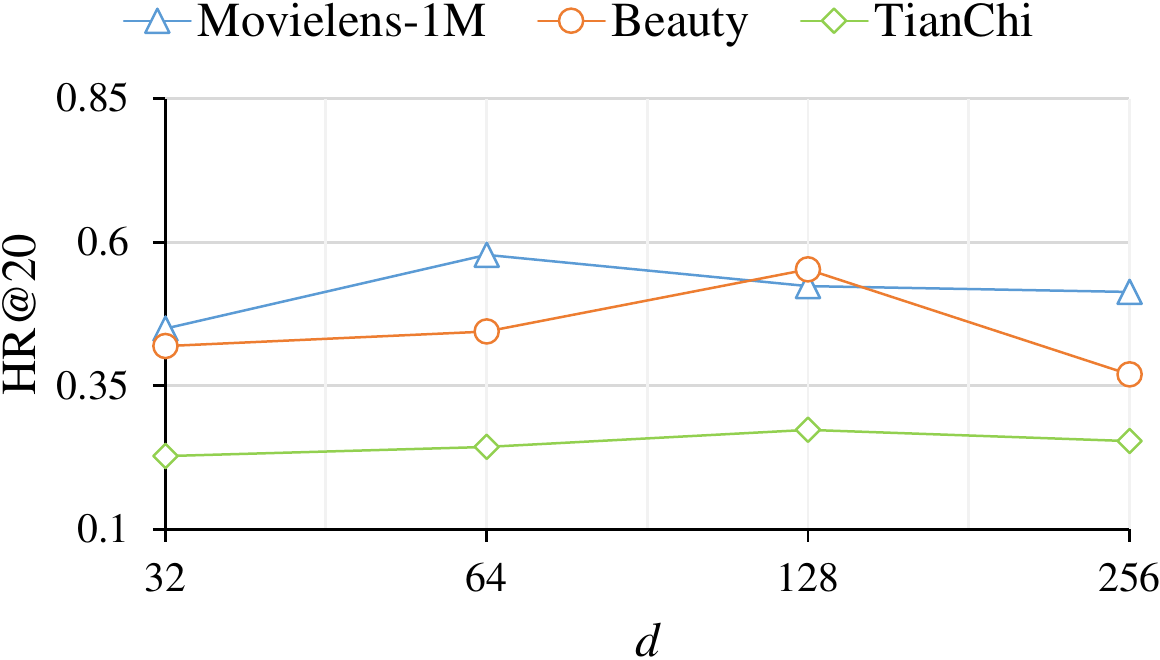}
\end{minipage}%
}
\subfigure[]{
\begin{minipage}[t]{0.3\textwidth}
\centering
\includegraphics[scale=0.45]{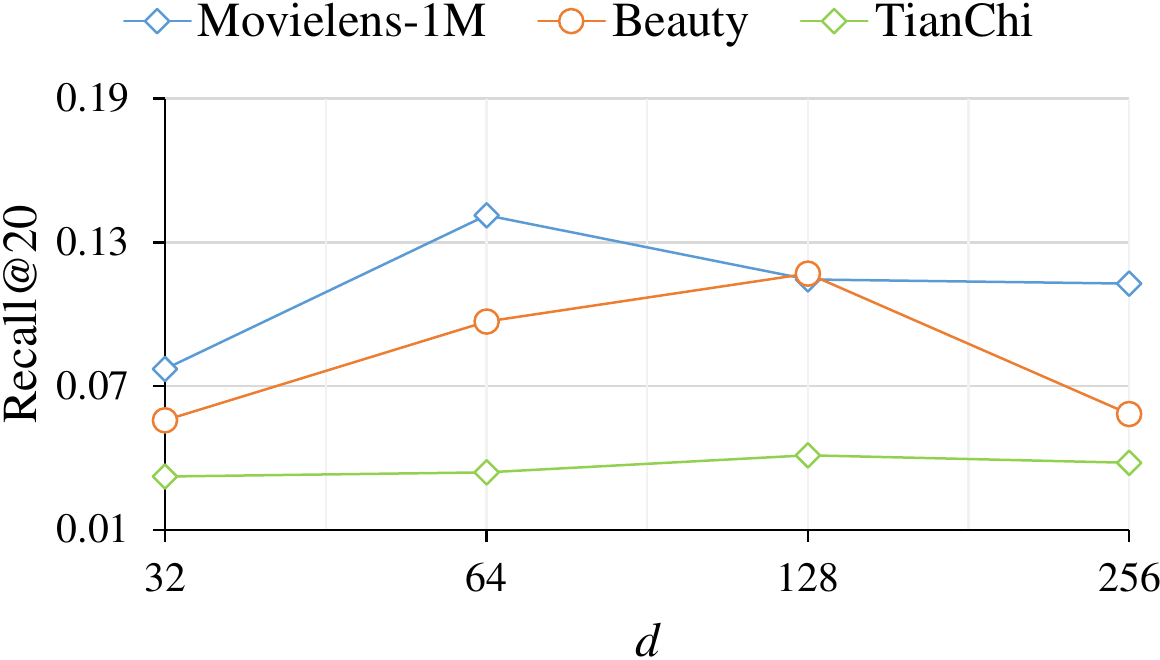}
\end{minipage}
}
\subfigure[]{
\begin{minipage}[t]{0.3\textwidth}
\centering
\includegraphics[scale=0.45]{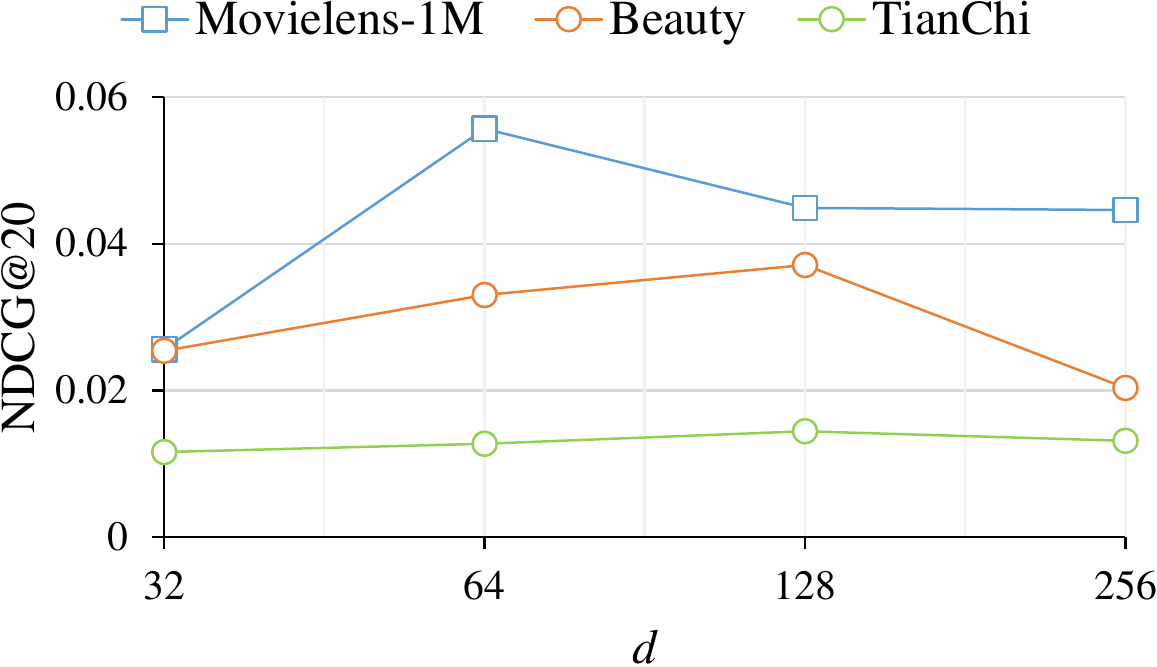}
\end{minipage}
}
\caption{Tuning of the embedding size $d$.}
\label{fig:figure-d}
\end{figure*}

\subsection{Case Study (RQ3)}

Now we conduct a case study to illustrate TIARec's ability to capture temporary interest and make recommendations with balance between the general preference and temporary interest of a user. For this purpose, we sample one user with ID "264" from Movielens-1M. Figure \ref{fig:preference} uses a word cloud to visualize the tag distribution of the movies watched by the sample user in 2002, where the size of a tag word is proportional to the frequency that the sample user watched the movies with that tag. From Figure \ref{fig:preference} we can see that in 2002, the sample user generally preferred drama and comedy, and seldom watched movies of other types. However, in October and November the sample user watched some war movies for some unknown reason, as marked with a red circle in Figure \ref{fig:preference}. We randomly choose one war movie the sample user watched in November as the ground-truth for test, and check the top-20 movies recommended by TIARec and the baselines, whose tag distributions are shown in Figures \ref{fig:TIARec} to \ref{fig:APR}, respectively. 

From Figure \ref{fig:TIARec} we can see that the drama movies and war movies account for the vast majority of the top-20 list recommended by TIARec. In fact, 6 war movies together with 10 drama movies are recommended in the top-20 list, including 4 movies of both drama and war, which illustrates that TIARec is able to capture the sample user's temporary interest in war movie and can balance it with her general preference to drama when making recommendations. We also note that as another general preference, the comedy movies are much less than the drama ones. We argue that this is because as war movie is more compatible with drama movie, quite a few war movies are also tagged with drama. To make an easy tradeoff between the temporary interest and the general preference of the sample user in this case, TIARec tends to combine war movies with drama rather than comedy ones.

From Figure \ref{fig:TIARec-C} we can see that without the help of the auxiliary classifier agent, TIARec-C pays most attention to the sample user's general preference to comedy movies and can hardly recommend war movies. Compared with Figure \ref{fig:TIARec}, Figure \ref{fig:TIARec-C} shows that it is the classifier agent that offers TIARec the ability to learn a user's temporary interest from her atypical behaviors.

Figures \ref{fig:PDQ} to \ref{fig:APR} show that similar to TIARec-C, the sample user's general preference to drama or comedy dominates the recommendations made by the baselines. In fact, none of them recommends more than 2 war movies, which indicates that the baselines can hardly capture the sample user's temporary interest in war movie from the recent watching of the war movies, and excessively rely on the general preference of users when making recommendations. Compared with Figure \ref{fig:TIARec}, such results verify the superiority of TIARec in the temporal interest learning for robust recommendation.

\subsection{Robustness Study (RQ4)}
Now we investigate the robustness of TIARec by comparing it with the baseline methods on the three datasets, in terms of the metric HR@10. For this purpose, in each testing set, with respect to a given ratio (noise level) we randomly change some consecutive five interactions to the items of the same category whose frequency is less than the 20\% of the average, to simulate the atypical behaviors that can be regarded as temporary interest. Figure \ref{fig:robust} shows the results at noise levels 0.1, 0.2, 0.4, and 0.8. From Figure \ref{fig:robust} we can see that on each dataset the performances of all methods decline as noise level rises. However, TIARec still consistently outperforms the baseline methods at different noise levels, which again shows the advantage of TIARec in capturing temporary interest of a user. In particular, it is worthy to note that when the noise level is 0.8, the interest of almost every user becomes very unstable and changes quite frequently (every five interactions), which results in a drastic deterioration of the performances of all the methods. However, even in such extreme case, the HR@10 of TIARec still remains greater than 0.1 which is remarkably higher than that of the baseline methods.

\subsection{Hyper-parameter Tuning (RQ5)}
At last we tune the embedding size $d$ on validation sets, of which the results are shown in Figure \ref{fig:figure-d}. We can observe that the performance curves first increase and then decrease when we enlarge the embedding size $d$. The performance reaches its maximum at $d=64$ on Movielens-1M and at $d=128$ on both Beauty and Tianchi. A larger $d$ means more model parameters needed to be learned, which may lead to the problem of overfitting.

\section{RELATED WORK}

In this section, we briefly review two domains of the works mostly related to our study, including robust recommendation and RL-based recommendation.

\subsection{Robust Recommendation}
The goal of robust recommendation is to reduce the impact of atypical interactions on user preference learning as much as possible to improve the robustness of the models. The existing methods for robust recommendation can be roughly divided into two classes. One class of the methods employs auto-encoder based techniques to generate robust embeddings for users and items \cite{CDAE,Li2017Collaborative,shenbin2020recvae}. For example, Wu et al. propose the model CDAE \cite{CDAE} which first injects random drop-out noise to training data and trains an denoising auto-encoder \cite{Vincent2008Extracting} on intentionally corrupted input with the objective of minimizing reconstruction errors. Li et al. \cite{Li2017Collaborative} propose a collaborative variational auto-encoder (CVAE) for robust recommendation of multimedia, and Shenbin et al. \cite{shenbin2020recvae} propose a Recommender VAE (RecVAE) model with that can be trained with corrupted implicit user-item interaction vectors. The other class of the robust recommendation methods introduces adversarial noise as well as adversarial training to improve the model robustness \cite{APR,Tang2018Adversarial,wang2017irgan,Yuan2019Adversarial}. For example, He et al. propose the model APR \cite{APR} which enhances the pairwise ranking method BPR \cite{Rendle2012BPR} with an adversarial training. Similarly, Wang et al. \cite{wang2017irgan}, Tang et al. \cite{Tang2018Adversarial}, and Yuan et al. \cite{Yuan2019Adversarial} propose to add perturbations to the input data and use adversarial learning to strengthen the robustness of recommendation model. As we have mentioned, the existing robust recommendations methods often neglect the meaningful signal carried by the atypical interactions, from which our TIARec can learn users' temporary interest for making better recommendations.

\subsection{RL-based Recommendation}
Recently, RL-based recommendation approaches have attracted more attentions from scholars. RL-based models aim to learn an optimal strategy to maximize the long terms rewards. RL-based models can be divided into three categories, the policy-based methods \cite{DDPG,zhao2017deep,Zhao2018,chen2018large-scale,liu2018deep,chen2019top-k,zhao2019model}, the value-based methods\cite{DEERs,zheng2018drn,zou2020pseudo,Liu2020}, and the actor-critic based methods \cite{zhang2017dynamic,gui2019mention,feng2018learning,he2020learning,zhao2020,Xin2020}. Chen et al. \cite{chen2018large-scale} propose to use a balanced hierarchical clustering tree to tackle the large action space problem. Chen et al. \cite{chen2019top-k} propose to use an off-policy framework to solve the data bias problem. Liu et al. \cite{liu2018deep} model the interactions between users and items to represent users' state. Zhao et al. \cite{DEERs} take the negative feedback and positive feedback into consideration to represent users' state. Zheng et al. \cite{zheng2018drn} utilize two Q-network to balance the correlation between exploit and explore. Zou et al. \cite{zou2020pseudo} construct a simulator to model users' feedback to handle the selection bias of logged data. Zhou et al. \cite{Zhou2020} propose to obtain item representations through a graph convolution network and obtain users' state embedding vectors through a recurrent neural network, and then trains the recommendation strategy with the Q-leaning algorithm. He et al. \cite{he2020learning} and Feng et al. \cite{feng2018learning} treat each agent as a scenario and use a multi-agent framework to improve the total performance of all scenarios. Zhang et al. \cite{zhang2017dynamic} apply a multi-agent reinforcement learning method to coauthor network analysis for the dynamic collaboration recommendation. Zhao et al. \cite{zhao2017deep} propose an actor-critic based reinforcement learning method for list-wise recommendation and introduce a simulator of online user-agent interacting environment to get more interactions. Zhao et al. \cite{zhao2019model} also propose a multi-agent reinforcement learning model to capture the sequential correlation among different scenarios. The existing RL-based recommendation methods equally treat user-item interactions as normal, which offer little robustness against atypical interactions and cannot capture users' temporary interest too.

\section{CONCLUSIONS}
In this paper, we propose a novel reinforcement learning based model for robust recommendation, called Temporary Interest Aware Recommendation (TIARec), which can learn users' temporary interest from their atypical interactions. TIARec contains a recommender agent and an auxiliary classifier agent, which are jointly trained with the objective of maximizing the expected cumulative return of the recommendations made by the recommender agent. In particular, during the joint training, the classifier agent can judge whether the interaction with a recommended item is atypical and transfer the knowledge about the temporary interest learning to the recommender agent, which finally enable the recommender agent to alone make robust recommendations that balance the general preference and temporary interest of users. The experiments conducted on real world datasets show that due to the ability to capture users' temporary interest, TIARec achieves a significant improvement in recommendation performance compared with the state-of-the-art baselines.

\section*{Acknowledgment}
This work is supported by National Natural Science Foundation of China under grant 61972270, and in part by NSF under grants III-1763325, III-1909323,  III-2106758, and SaTC-1930941.

\bibliographystyle{abbrv}
\bibliography{TIARec_TKDE}

\begin{thebibliography}{10}

\bibitem{chen2018large-scale}
H.~Chen, X.~Dai, H.~Cai, W.~Zhang, X.~Wang, R.~Tang, Y.~Zhang, and Y.~Yu.
\newblock Large-scale interactive recommendation with tree-structured policy
  gradient.
\newblock In {\em AAAI}, 2019.

\bibitem{chen2019top-k}
M.~Chen, A.~Beutel, P.~Covington, S.~Jain, F.~Belletti, and E.~H. Chi.
\newblock Top-k off-policy correction for a reinforce recommender system.
\newblock In {\em RecSys}, 2019.

\bibitem{Deldjoo2020}
Y.~Deldjoo, T.~Di~Noia, E.~Di~Sciascio, and F.~A. Merra.
\newblock How dataset characteristics affect the robustness of collaborative
  recommendation models.
\newblock In {\em SIGIR}, 2020.

\bibitem{DDPG}
G.~Dulacarnold, R.~Evans, P.~Sunehag, and B.~Coppin.
\newblock Deep reinforcement learning in large discrete action spaces.
\newblock {\em CoRR}, 2015.

\bibitem{feng2018learning}
J.~Feng, H.~Li, M.~Huang, S.~Liu, W.~Ou, Z.~Wang, and X.~Zhu.
\newblock Learning to collaborate: Multi-scenario ranking via multi-agent
  reinforcement learning.
\newblock In {\em WWW}, 2018.

\bibitem{gui2019mention}
T.~Gui, P.~Liu, Q.~Zhang, L.~Zhu, M.~Peng, Y.~Zhou, and X.~Huang.
\newblock Mention recommendation in twitter with cooperative multi-agent
  reinforcement learning.
\newblock In {\em SIGIR}, 2019.

\bibitem{he2020learning}
X.~He, B.~An, Y.~Li, H.~Chen, R.~Wang, X.~Wang, R.~Yu, X.~Li, and Z.~Wang.
\newblock Learning to collaborate in multi-module recommendation via
  multi-agent reinforcement learning without communication.
\newblock In {\em RecSys}, 2020.

\bibitem{APR}
X.~He, Z.~He, X.~Du, and T.-S. Chua.
\newblock Adversarial personalized ranking for recommendation.
\newblock In {\em SIGIR}, pages 355--364, 2018.

\bibitem{ie2019slateq}
E.~Ie, V.~Jain, J.~Wang, S.~Narvekar, R.~Agarwal, R.~Wu, H.-T. Cheng,
  T.~Chandra, and C.~Boutilier.
\newblock Slateq: A tractable decomposition for reinforcement learning with
  recommendation sets.
\newblock In {\em IJCAI}, 2019.

\bibitem{Li2017Collaborative}
X.~Li and J.~She.
\newblock Collaborative variational autoencoder for recommender systems.
\newblock In {\em KDD}, 2017.

\bibitem{Liu2020}
F.~Liu, H.~Guo, X.~Li, R.~Tang, Y.~Ye, and X.~He.
\newblock End-to-end deep reinforcement learning based recommendation with
  supervised embedding.
\newblock In {\em WSDM}, 2020.

\bibitem{liu2018deep}
F.~Liu, R.~Tang, X.~Li, Y.~Ye, H.~Chen, H.~Guo, and Y.~Zhang.
\newblock Deep reinforcement learning based recommendation with explicit
  user-item interactions modeling.
\newblock {\em arXiv:1810.12027}, 2018.

\bibitem{Lowe2017}
R.~Lowe, Y.~WU, A.~Tamar, J.~Harb, O.~Pieter~Abbeel, and I.~Mordatch.
\newblock Multi-agent actor-critic for mixed cooperative-competitive
  environments.
\newblock In {\em Advances in Neural Information Processing Systems}, 2017.

\bibitem{lumbantoruan2019cars}
R.~Lumbantoruan, X.~Zhou, Y.~Ren, and L.~Chen.
\newblock I-cars: an interactive context-aware recommender system.
\newblock In {\em ICDM}, 2019.

\bibitem{DQN}
V.~Mnih, K.~Kavukcuoglu, D.~Silver, A.~Graves, I.~Antonoglou, D.~Wierstra, and
  M.~Riedmiller.
\newblock Playing atari with deep reinforcement learning.
\newblock {\em arXiv:1312.5602}, 2013.

\bibitem{Rendle2012BPR}
S.~Rendle, C.~Freudenthaler, Z.~Gantner, and L.~Schmidt-Thieme.
\newblock Bpr: Bayesian personalized ranking from implicit feedback.
\newblock In {\em UAI}, 2009.

\bibitem{shang2019environment}
W.~Shang, Y.~Yu, Q.~Li, Z.~Qin, Y.~Meng, and J.~Ye.
\newblock Environment reconstruction with hidden confounders for reinforcement
  learning based recommendation.
\newblock In {\em KDD}, 2019.

\bibitem{shenbin2020recvae}
I.~Shenbin, A.~Alekseev, E.~Tutubalina, V.~Malykh, and S.~I. Nikolenko.
\newblock Recvae: A new variational autoencoder for top-n recommendations with
  implicit feedback.
\newblock In {\em WSDM}, 2020.

\bibitem{Tang2018Adversarial}
J.~Tang, X.~Du, X.~He, F.~Yuan, Q.~Tian, and T.~Chua.
\newblock Adversarial training towards robust multimedia recommender system.
\newblock {\em TKDE}, 2020.

\bibitem{Vincent2008Extracting}
P.~Vincent, H.~Larochelle, Y.~Bengio, and P.-A. Manzagol.
\newblock Extracting and composing robust features with denoising autoencoders.
\newblock In {\em ICML}, 2008.

\bibitem{wang2017irgan}
J.~Wang, L.~Yu, W.~Zhang, Y.~Gong, Y.~Xu, B.~Wang, P.~Zhang, and D.~Zhang.
\newblock Irgan: A minimax game for unifying generative and discriminative
  information retrieval models.
\newblock In {\em SIGIR}, 2017.

\bibitem{Yao2016Collaborative}
Y.~Wu, C.~DuBois, A.~X. Zheng, and M.~Ester.
\newblock Collaborative denoising auto-encoders for top-n recommender systems.
\newblock In {\em WSDM}, 2016.

\bibitem{CDAE}
Y.~Wu, C.~DuBois, A.~X. Zheng, and M.~Ester.
\newblock Collaborative denoising auto-encoders for top-n recommender systems.
\newblock In {\em WSDM}, 2016.

\bibitem{Xin2020}
X.~Xin, A.~Karatzoglou, I.~Arapakis, and J.~M. Jose.
\newblock Self-supervised reinforcement learning for recommender systems.
\newblock In {\em SIGIR}, 2020.

\bibitem{Yuan2019Adversarial}
F.~Yuan, L.~Yao, and B.~Benatallah.
\newblock Adversarial collaborative neural network for robust recommendation.
\newblock In {\em SIGIR}, 2019.

\bibitem{zhang2019hierarchical}
J.~Zhang, B.~Hao, B.~Chen, C.~Li, H.~Chen, and J.~Sun.
\newblock Hierarchical reinforcement learning for course recommendation in
  moocs.
\newblock In {\em AAAI}, pages 435--442, 2019.

\bibitem{Zhang2020}
S.~Zhang, H.~Yin, T.~Chen, Q.~V.~N. Hung, Z.~Huang, and L.~Cui.
\newblock Gcn-based user representation learning for unifying robust
  recommendation and fraudster detection.
\newblock In {\em SIGIR}, 2020.

\bibitem{zhang2017dynamic}
Y.~Zhang, C.~Zhang, and X.~Liu.
\newblock Dynamic scholarly collaborator recommendation via competitive
  multi-agent reinforcement learning.
\newblock In {\em RecSys}, 2017.

\bibitem{zhao2020}
D.~Zhao, L.~Zhang, B.~Zhang, L.~Zheng, Y.~Bao, and W.~Yan.
\newblock Mahrl: Multi-goals abstraction based deep hierarchical reinforcement
  learning for recommendations.
\newblock In {\em SIGIR}, 2020.

\bibitem{zhao2019model}
X.~Zhao, L.~Xia, D.~Yin, and J.~Tang.
\newblock Model-based reinforcement learning for whole-chain recommendations.
\newblock {\em arXiv:1902.03987}, 2019.

\bibitem{Zhao2018}
X.~Zhao, L.~Xia, L.~Zhang, Z.~Ding, D.~Yin, and J.~Tang.
\newblock Deep reinforcement learning for page-wise recommendations.
\newblock In {\em RecSys}, 2018.

\bibitem{DEERs}
X.~Zhao, L.~Zhang, Z.~Ding, L.~Xia, J.~Tang, and D.~Yin.
\newblock Recommendations with negative feedback via pairwise deep
  reinforcement learning.
\newblock In {\em KDD}, pages 1040--1048, 2018.

\bibitem{zhao2017deep}
X.~Zhao, L.~Zhang, L.~Xia, Z.~Ding, D.~Yin, and J.~Tang.
\newblock Deep reinforcement learning for list-wise recommendations.
\newblock In {\em arXiv:1801.00209}, 2017.

\bibitem{zheng2018drn}
G.~Zheng, F.~Zhang, Z.~Zheng, Y.~Xiang, N.~J. Yuan, X.~Xie, and Z.~Li.
\newblock Drn: A deep reinforcement learning framework for news recommendation.
\newblock In {\em WWW}, 2018.

\bibitem{Zhou2020}
S.~Zhou, X.~Dai, H.~Chen, W.~Zhang, K.~Ren, R.~Tang, X.~He, and Y.~Yu.
\newblock Interactive recommender system via knowledge graph-enhanced
  reinforcement learning.
\newblock In {\em SIGIR}, 2020.

\bibitem{zou2019reinforcement}
L.~Zou, L.~Xia, Z.~Ding, J.~Song, W.~Liu, and D.~Yin.
\newblock Reinforcement learning to optimize long-term user engagement in
  recommender systems.
\newblock In {\em KDD}, 2019.

\bibitem{zou2020pseudo}
L.~Zou, L.~Xia, P.~Du, Z.~Zhang, T.~Bai, W.~Liu, J.-Y. Nie, and D.~Yin.
\newblock Pseudo dyna-q: A reinforcement learning framework for interactive
  recommendation.
\newblock In {\em WSDM}, 2020.

\end{thebibliography}

\begin{IEEEbiography}[{\includegraphics[width=1in,height=1.2in,clip,keepaspectratio]{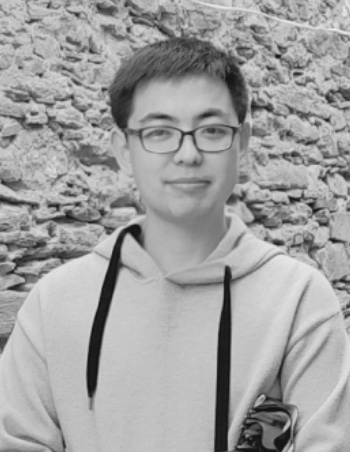}}]{Ziwen Du} obtained his bachelor's degree from the School of Computer Science, Sichuan University, China, in 2018. He is now pursuing the master's degree in the School of Computer Science, Sichuan University, China. His research interests include data mining and recommender systems. 
\end{IEEEbiography}

\begin{IEEEbiography}[{\includegraphics[width=1in,height=1.25in,clip,keepaspectratio]{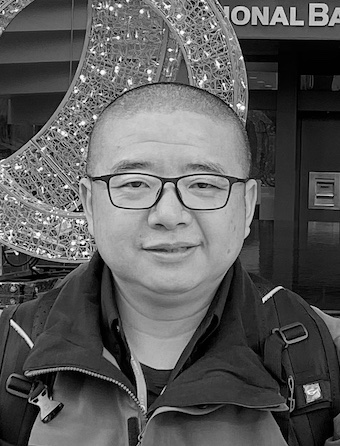}}]{Ning Yang} is an associate professor at Sichuan University, China. He obtained his PhD degree in Computer Science from Sichuan University in 2010. His research interests include recommender systems and social media mining.
\end{IEEEbiography}

\begin{IEEEbiography}[{\includegraphics[width=1in,height=1.25in,clip,keepaspectratio]{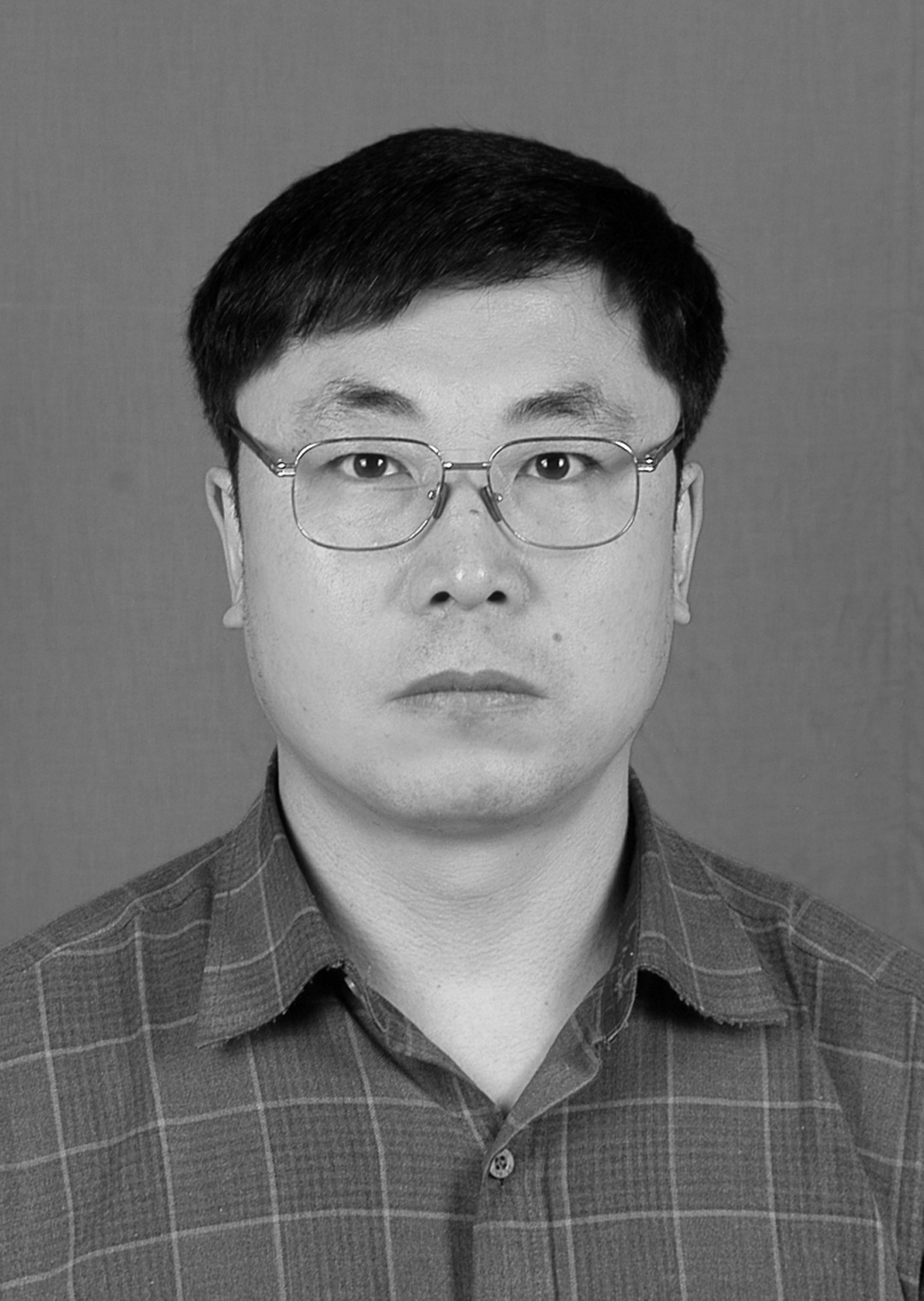}}]{Zhonghua Yu} is an associate professor at Sichuan University, China. He obtained his PhD degree in Computer Science from Belarusian State University in 1995. He was a post-doctor of the University of Tokyo from 2001 to 2003, with the supervision of Junichi Tsujii. His research interests include natural language processing and machine learning.
\end{IEEEbiography}

\begin{IEEEbiography}[{\includegraphics[width=1in,height=1.25in,clip,keepaspectratio]{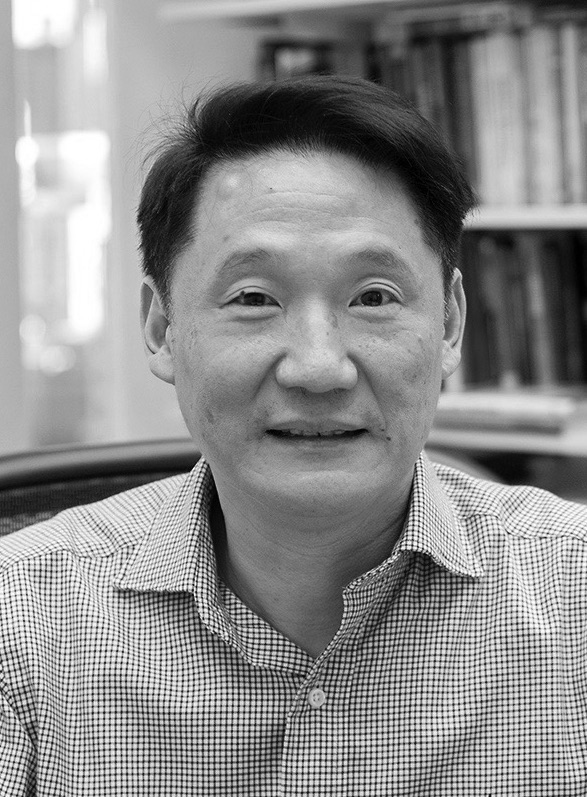}}]{Philip S. Yu} received the PhD degree in electrical engineering from Stanford University. He is a distinguished professor in computer science at the University of Illinois at Chicago and is also the Wexler chair in information technology. His research interests include big data, data mining, and social computing. He is a fellow of the ACM and the IEEE.
\end{IEEEbiography}

\end{document}